\documentclass[11pt,a4paper]{article}

\usepackage[a4paper,margin=1in]{geometry}
\usepackage{amsmath,amssymb,amsfonts,amsthm}
\usepackage{graphicx}
\usepackage{float}
\usepackage{multirow}
\usepackage{adjustbox}
\usepackage{array}
\usepackage{dsfont}
\usepackage[numbers]{natbib}
\usepackage{hyperref}
\usepackage{authblk}
 \usepackage{booktabs}
\usepackage{multirow}
\usepackage{adjustbox}
\usepackage{mathtools}

\newcommand\BibTeX{{\rmfamily B\kern-.05em \textsc{i\kern-.025em b}\kern-.08em
T\kern-.1667em\lower.7ex\hbox{E}\kern-.125emX}}

\begin{document}

\newcommand{\indep}{\mathrel{\text{\scalebox{1.07}{$\perp\mkern-10mu\perp$}}}}

\title{Built-in Selection Bias in Proportional Hazards Models with Omitted Covariates: Simulation Evidence and Alternative Approaches}

\author[1]{Ayoub Bifenzi}
\author[1]{Hélène Jacqmin-Gadda\thanks{Corresponding author: Hélène Jacqmin-Gadda, Centre Inserm Bordeaux Population Health U1219, Université de Bordeaux, 146 rue Léo Saignat CS61292, 33076 Bordeaux CEDEX, France. Email: \href{mailto:helene.jacqmin-gadda@u-bordeaux.fr}{helene.jacqmin-gadda@u-bordeaux.fr}}}

\affil[1]{Univ. Bordeaux, INSERM, Bordeaux Population Health, U1219, France}

\date{}

\maketitle

\begin{abstract}

In time-to-event analysis, the hazard ratio (HR) derived from the Cox proportional hazards (PH) model is the most commonly used and widely reported measure for assessing treatment effects. However, hazard ratios are non-collapsible due to their inherent conditioning on survival up to each time point. As a result, they are subject to built-in selection bias in the presence of unmeasured heterogeneity arising from omitted important covariates, even when these covariates are independent of the main exposure at baseline, as is the case in randomized controlled trials. This article aims to provide an overview of key findings from the literature on how unobserved heterogeneity, due to omitted covariates that affect the outcome, can bias the estimation of the treatment hazard ratio in standard proportional hazards models, even in randomized trials where treatment is assigned independently of such covariates. Through simulations, we evaluate the extent of bias in the semi-parametric Cox PH model and parametric PH model under various scenarios of unmeasured heterogeneity. We then compare these standard models to alternative approaches that either account for this issue or are considered robust to it. These alternatives include the hazard ratio estimated from frailty models, regression parameters from an Accelerated Failure Time (AFT) model, and survival differences between treatment groups estimated nonparametrically using Kaplan-Meier curves or based on a Cox model with time-dependent effect of the exposure.  We illustrate the practical relevance of the explored alternatives through a real data application to a randomized controlled trial from the Radiation Therapy Oncology Group (RTOG 9202).

\end{abstract}


\section{Introduction}\label{sec:intro}

When analyzing time-to-event outcomes, the goal is often to understand how an individual's risk of experiencing an event changes depending on the exposure to a treatment or intervention. A common approach is the Cox proportional hazards model (PHM)\cite{cox1972regression}, which estimates a time-constant hazard ratio: the relative instantaneous risk of the event occurring in the exposed versus unexposed group, assumed to remain constant over time.

However, this approach has been extensively criticized, particularly following the argument of Hern{\'a}n\cite{hernan2010hazards}  that hazard ratios, even those obtained from randomized controlled trials, can be misleading for causal interpretation. 
The first concern is that the proportional hazards (PH) assumption is often violated in practice, as treatment effects are rarely constant over time. More importantly, even when the treatment effect is truly constant, hazard ratios remain vulnerable to built-in selection bias. In most time-to-event studies, not all relevant covariates are observed: some may be unmeasured, unknown, or not suspected to influence the outcome. The omission of such variables induces a dynamic selection process over the course of follow-up. This selection process is inherent to the definition of the hazard function, which conditions on survival up to the current time. Consequently, the hazard ratio is estimated on a changing subset of individuals who have not yet experienced the event. Over time, this risk set becomes progressively more selected: individuals at higher risk tend to fail earlier and are removed from the risk set, while less susceptible individuals remain. This selection does not occur symmetrically across treatment arms due to their different risks, resulting in differential depletion over time. As a consequence, the distribution of unmeasured covariates, balanced at baseline through randomization, can become imbalanced over time. When such covariates are not accounted for in the model, this imbalance introduces structural bias into the estimated hazard ratio. Specifically, conditioning on survival induces a spurious association between treatment and unmeasured covariates that were independent at baseline due to randomization, a phenomenon known in causal inference as the collider effect\cite{greenland2003quantifying}. This phenomenon occurs whenever we condition on a variable that is causally influenced by two or more other variables, as is the case with survival, which depends jointly on both treatment and unmeasured covariates. As a consequence, even in randomized trials, hazard ratios may reflect not only the causal effect of the treatment but also a form of selection bias intrinsic to the hazard ratio as a measure of association.

To further illustrate the bias induced by the collider effect, consider the example of studying risk factors for mortality, as represented by the following causal diagram:

\begin{figure*}[h]
 \centering
 \includegraphics[width=5cm]{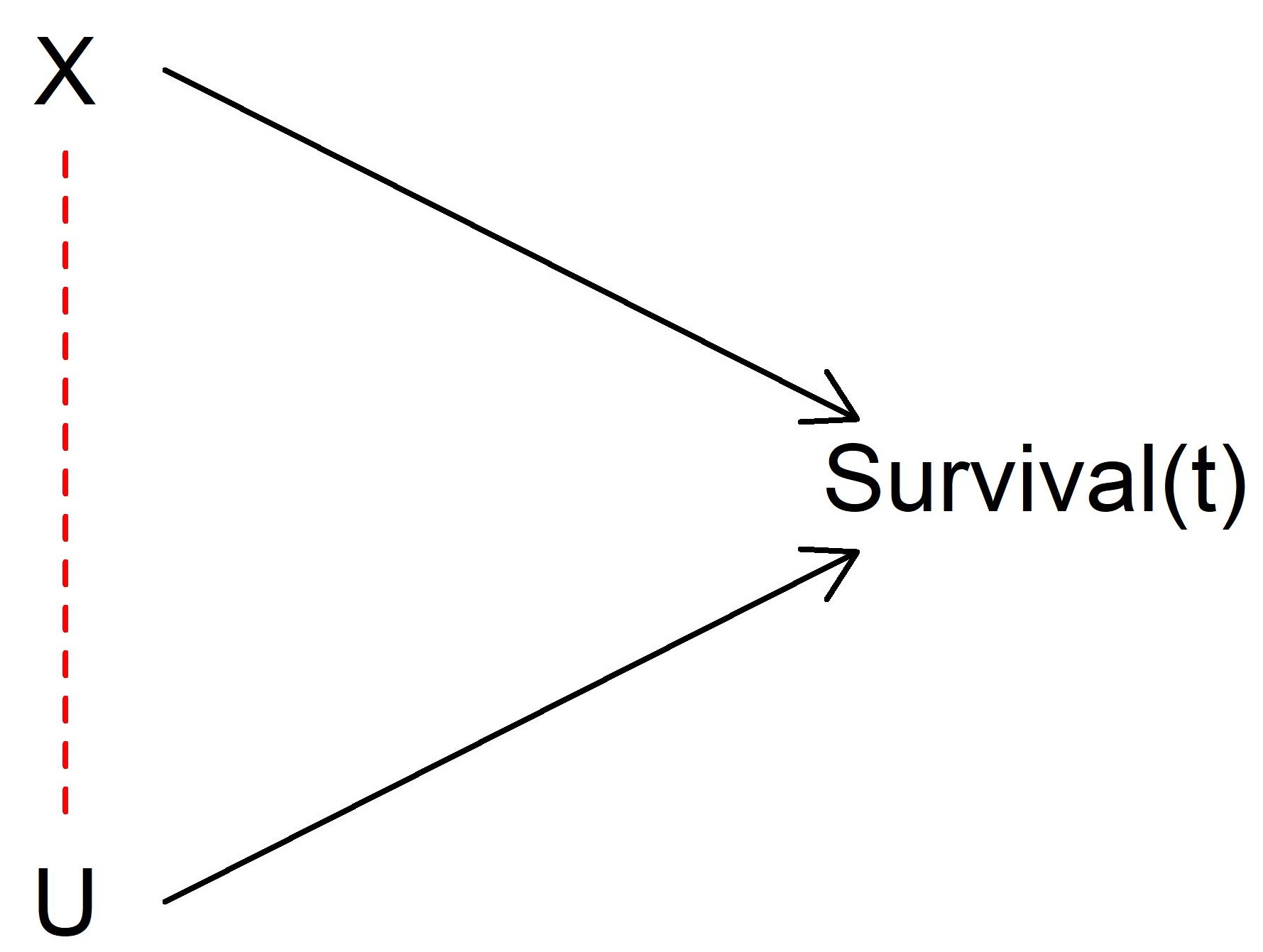}
 \caption{Directed acyclic graph showing the relationships between variables $X$, $U$, and survival at time $t$.}
 \label{fig1}
\end{figure*}

Here, $X$ denotes a binary treatment (1 for treated, 0 for untreated) that influences the risk of death. 
$U$ represents the "unmeasured" or "unobserved" factor that also affects the risk of death and is assumed to be independent of $X$ at baseline. The time to death is denoted by $T$.

The hazard function at time 
$t$, conditional on $X$ and $U$ is defined as:

\begin{equation*}
\displaystyle
\lambda(t \mid X, U) = \lim_{\Delta t \to 0^+} \frac{\mathbb{P}(t \le T < t + \Delta t \mid \boldsymbol{T > t}, X, U)}{\Delta t}
\end{equation*}

When we compute the hazard function at time $t$, we implicitly condition on individuals having survived up to time $t$. In the causal diagram, survival up to time $t$ acts as a collider variable, as it is influenced by both the treatment $X$ and $U$. Conditioning on it can induce a spurious statistical association between  $X$ and $U$, even if these variables were independent at baseline. That is, conditioning on survival breaks the balance in the distribution of $U$ across treatment groups that is initially ensured by randomization. In other words, $\boldsymbol{(X \not\perp U)\mid T>t}$ in general. As a result, hazard ratio estimates obtained from models that do not adjust for $U$, and which are inherently conditional on survival, can reflect not only the true causal effect of treatment, but also the induced imbalance in $U$ over time.

\begin{figure*}
 \centering
 \includegraphics[width=15.8cm]{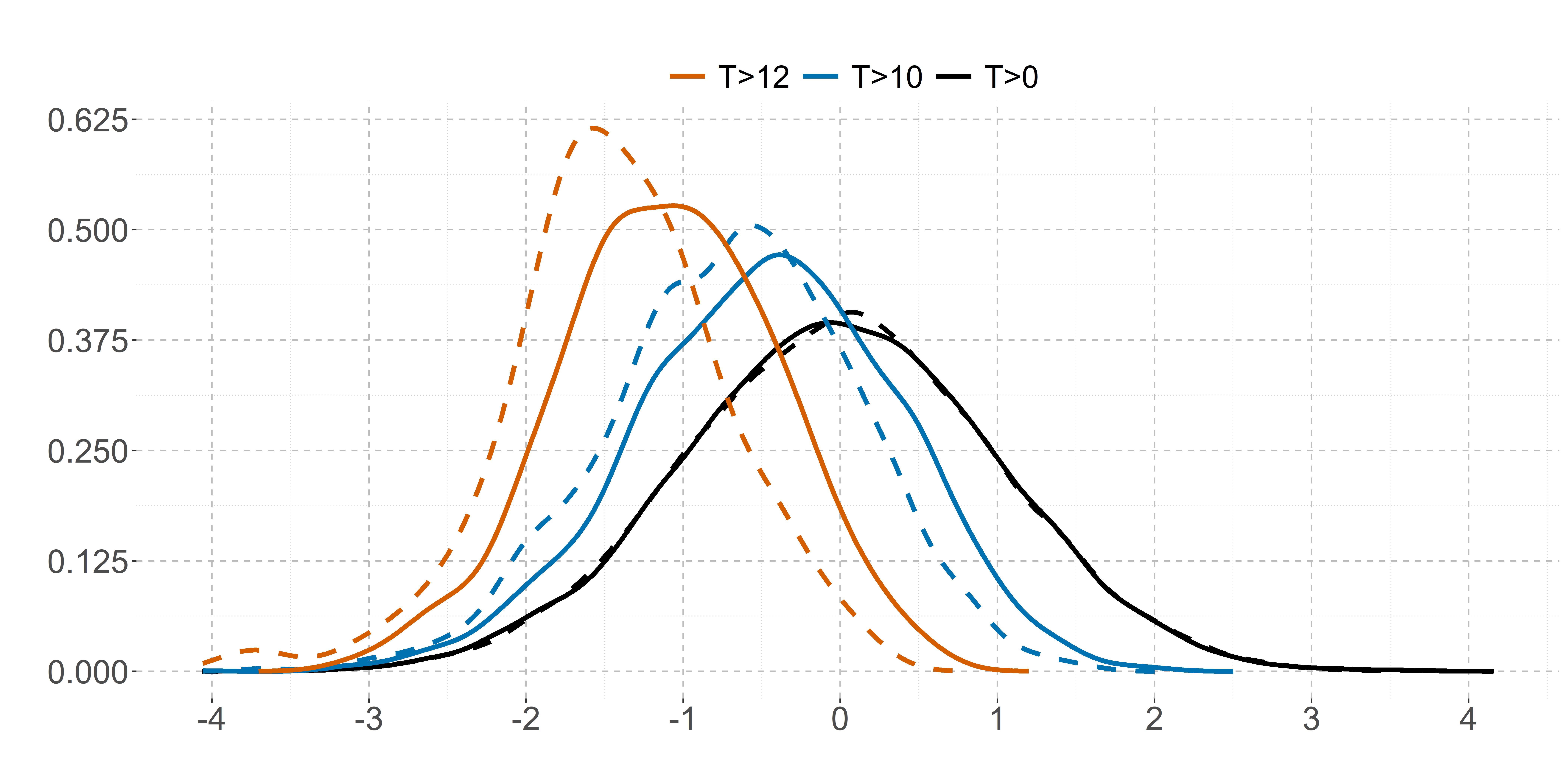}
 \caption{Estimated density of $U$ among individuals at risk at different time points, accross the two groups of treatment: treated group ($X=1$, solid lines) and non treated group ($X=0$, dashed lines).}
 \label{fig2}
\end{figure*}

To illustrate this mechanism more concretely, we simulated a dataset of size $n=40000$, where event times were generated from a PHM with Weibull baseline hazard, and  two explanatory variables $X \sim \mathcal{B}er(0.5)$ and $U \sim \mathcal{N}(0,1)$, having log-hazard ratios of $-0.6$ and $1$, respectively, using the cumulative hazard inversion method\cite{brilleman2021simulating}. The censoring times were drawn from a Weibull distribution, right truncated at $t=12.60$.
Figure~\ref{fig2} presents the estimated densities of $U$ given $X$ among individuals at risk at three distinct time points. At $T=0$, the densities in the two treatment groups coincide, as expected, since no selection has occurred yet, and $X$ and $U$ are independent at baseline. At $T=10$, the density of $U$ among subjects at risk in the group with $X=0$ is slightly shifted to the left relative to that in the treated group. This reflects the fact that, in order to remain at risk up to this time, subjects with $X=0$ are more likely to have negative values of $U$. At $T=12$, the same pattern is observed, with the difference between the two densities being more pronounced, due to the increasing selection over time.

This article aims to provide a comprehensive and pedagogical overview of key theoretical results from the literature on how unobserved heterogeneity caused by omitting  covariates that influence the outcome biases the estimation of the treatment hazard ratio in a standard PHM, even under randomized trial settings where treatment assignment is independent of such covariates. We then review several alternative approaches that either account for or are robust to such omissions. These alternatives include the hazard ratio estimated from frailty models, regression parameters from an Accelerated Failure Time (AFT) model, and survival differences between treatment groups estimated either nonparametrically using Kaplan-Meier curves or based on a Cox regression model with a time dependent effect of the exposure. Then through simulations, we quantify the extent of this bias under various scenarios of unobserved heterogeneity and assess the robustness of the proposed alternatives.

We introduce frailty models as a natural extension of Cox model that incorporates subject-specific random effects to account for unobserved heterogeneity. These models can offer a correction for the built-in selection bias that arises from the omission of $U$. While the use of frailty models for this purpose has been suggested in previous literature, for example, in Balan and Putter\cite{balan2020tutorial}, their robustness in this context has not been evaluated through simulations. We further examine AFT models as an alternative to the Cox PH framework and provide simulations to assess their robustness. AFT models provide estimates that are not affected by the built-in selection bias and are collapsible, offering a more straightforward causal interpretation. While survival differences have been suggested as an alternative \cite{stensrud2025use}, to our knowledge, comprehensive simulations to assess their efficacy in this specific context have not yet been conducted. 

We then present an application to a randomized controlled trial from the Radiation Therapy Oncology Group (RTOG9202)\cite{lawton2017duration} to show the practical usefulness of the explored methods and to illustrate their use in applied settings.

\section{Built-in selection}
\label{sec:Builtinselection}

\subsection{Notations}

We will assume throughout the article that the time-to-event variable for a subject $i$, denoted by $T_i$ is governed by a hazard function following a PH structure, given by:

\begin{equation}
\label{eq:true model}
\displaystyle
\lambda_{C}(t_i \mid X_i,U_i)=\lambda_{0,C}(t_i)\exp(\beta_{C}X_i+\beta_U U_i) 
\end{equation}

Here, $\lambda_{0,C}(.)$ denotes the baseline hazard function, and $X_i$ and $U_i$ represent the values of the covariates $X$ and $U$ for subject $i$, that are assumed to be independent: $X_i \perp U_i$. The parameter $\beta_{C}$ corresponds to the log-hazard ratio for $X$ adjusted for $U$, while $\beta_{U}$ is the log-hazard ratio associated with the unmeasured covariate $U$.

The main concern is how the omission of the covariate $U$ impacts the
estimation of the hazard ratio for $X$ estimated from a PH
model. This unadjusted model takes the form:

\begin{equation}
\label{eq:unadjusted cox model}
\displaystyle
\lambda_{unadj}(t_i \mid X_i)=\lambda_{0,unadj}(t_i)\exp(\beta_{unadj}X_i) 
\end{equation}

Here, $\lambda_{0,unadj}(.)$ denotes the baseline hazard function, and $\beta_{unadj}$ represents the log-hazard ratio for $X$ when $U$ is omitted from the model. 

Specifically, we will discuss how $\beta_{unadj}$ deviates from the true conditional log-hazard ratio $\beta_{C}$ , illustrating the concept of non-collapsibility of hazard ratios. Then we will briefly review some  properties of the partial likelihood estimate of $\beta_{unadj}$ and discuss its interpretation and why it may be limited for drawing causal conclusions.

When there is no ambiguity, we will omit the subject-specific index $i$ for notational simplicity.

\subsection{Built-in selection leads to non collapsibility}

Following the definition of collapsibility given by Pearl \cite{pearl2009causality}, a measure of association (e.g., hazard ratios, odds ratios) is said to be \emph{collapsible} if, for a variable $U$ that is independent of the main exposure $X$, the marginal association measure for $X$ is equal to the average, over $U$, of the association measures for $X$ conditional on $U$. In many cases, the conditional association does not vary with the value of the omitted covariate $U$. In such cases, a measure of association is said to be \emph{strictly collapsible} if the association measure for $X$, adjusted for $U$, is identical to the marginal association for $X$. In this article, except when discussing survival differences, we do not distinguish between collapsibility and strict collapsibility, since all conditional measures considered do not depend on $U$.

The hazard ratio is not a collapsible measure of association, that is

\begin{align}
\label{eq:collaps1}
\mathbb{E}_U\left( 
    \frac{\lambda_{C}(t \mid X=1, U)}{\lambda_{C}(t \mid X=0, U)} 
\right) =  \exp(\beta_{C}) \notag \\
\neq \frac{\lambda_{M}(t \mid X=1)}{\lambda_{M}(t \mid X=0)} 
= HR_M(t)
\end{align}

where $\lambda_{M}(. \mid X)$ is the true marginal hazard function that derives from the true conditional model assumed in equation \eqref{eq:true model}, and $HR_M(.)$ denotes the true marginal hazard ratio. This true marginal hazard function, $\lambda_{M}(. \mid X)$, should not be confused with $\lambda_{unadj}(. \mid X)$, which is the hazard function under the unadjusted PH model defined in equation \eqref{eq:unadjusted cox model}.

Indeed, we have $\forall t>0$

\begin{align*}
    \lambda_M(t \mid X = x) 
    &= \lim_{\Delta t \to 0^+} \frac{\mathbb{P}(t \leq T < t+\Delta t \mid T > t, X = x)}{\Delta t} \\
    &= \lim_{\Delta t \to 0^+} \int \frac{\mathbb{P}(t \leq T < t+\Delta t \mid T > t, X = x, U = u)}{\Delta t} f_{U \mid T > t, X = x}(u) \, du \\
    &= \int \lambda_{C}(t \mid X = x, U = u) f_{U \mid T > t, X = x}(u) \, du \\
    &= \int \lambda_{0,C}(t) \exp(\beta_{C} x + \beta_U u) f_{U \mid T > t, X = x}(u) \, du \\
    &= \lambda_{0,C}(t)\exp(\beta_{C} x) \int \exp(\beta_U u) f_{U \mid T > t, X = x}(u) \, du 
\end{align*}

That is, $\forall t>0$

\begin{equation}
\label{eq:marginalhazard}
\lambda_M(t\mid X=x)=\lambda_{0,C}(t)\exp(\beta_{C}x)\mathbb{E}_{U\mid T>t,X=x}\left(\exp(\beta_U U)\right)
\end{equation}

where $\mathbb{E}_{U\mid T>t,X=x}\left(\exp(\beta_U U)\right)$ denotes the expectation of $\exp(\beta_U U)$ with respect to the distribution of $U$ conditionally on $T>t$ and $X=x$. The interchange of the limit and the integral from the second to the third line is rigorously established in Post et al.\cite{post2024built}.

It follows by taking the ratio of $\lambda_M(t\mid X=x)$ with $X=1$ and $X=0$ that:

\begin{equation}
\label{eq:collaps3}
\displaystyle
HR_M(t)=\exp(\beta_{C})\frac{\mathbb{E}_{U\mid T>t,X=1}\left(\exp(\beta_U U)\right)}{\mathbb{E}_{U\mid T>t,X=0}\left(\exp(\beta_U U)\right)}
\end{equation}

Furthermore, assuming that $\beta_{C} \neq 0$ and $\beta_{U} \neq 0$, we have (proof in ~\ref{sec:expectations-inequality})

\begin{equation}
\label{eq:expecineq}
\displaystyle
\mathbb{E}_{U\mid T>t,X=1}\left(\exp(\beta_U U)\right) \neq \mathbb{E}_{U\mid T>t,X=0}\left(\exp(\beta_U U)\right)
\end{equation}

Thus,

\begin{equation}
\displaystyle
HR_M(t)\neq \exp(\beta_{C})
\end{equation}

This shows that the hazard ratio is non-collapsible. Note that if $\beta_{C}=0$ or $\beta_U=0$, then $HR_M(t)=\exp(\beta_{C})$. This occurs because, in these cases, survival up to time $t$ no longer acts as a collider in Figure~\ref{fig1}. For example, if $\beta_{C}=0$, then $X$ has no effect on the time to event, and the arrow from $X$ to Survival$(t)$ disappears. Note also that equation \eqref{eq:marginalhazard} implies that the true marginal hazard function no longer follows, in general, a PHM structure. Furthermore, it follows from Equation~\eqref{eq:collaps3} together with the inequality~\eqref{eq:expecineq}, that $HR_M(t)$ is in fact a time-dependent function.

\subsection{Special case: Gamma frailty model}

In general, the true marginal hazard ratio $HR_M(t)$ does not have a closed-form expression because the expectations involved require intractable integrals. However, in some special cases, $HR_M(t)$ does admit an analytical form, for instance, when $\beta_U=1$ and $U$ follows a log-gamma distribution (or equivalently, when $\exp(U)$ follows a gamma distribution) with 
$\mathbb{E}(U)=1$ and $\mathrm{Var}(U)=\theta$. 
Other distributions for $U$ may also lead to closed-form expressions for $HR_M(t)$ as discussed in Balan and Putter\cite{balan2020tutorial}, but we focus here on the log-gamma distribution as an example.

In this case, model~\eqref{eq:true model} can be rewritten as:

\begin{equation}
\label{eq:true model 2}
\lambda_{C}(t \mid X,U)=\lambda_{0,C}(t)\exp(U)\exp(\beta_{C}X)
\end{equation}

This form is known as the gamma frailty model. As noted in Balan and Putter\cite{balan2020tutorial} and Aalen et al.\cite{aalen2015does}, and shown in   ~\ref{sec:proof-analytic-hazard}, the marginal hazard function is, for all $t>0$ given by:

\begin{equation}
\label{eq:analytic hazard function}
\lambda_M(t|X = x) = \frac{\lambda_{0,C}(t) \exp(\beta_{C}x)}{1 + \theta A_{0,C}(t) \exp(\beta_{C}x)}
\end{equation}

with $A_{0,C}(t) = \int_0^t \lambda_{0,C}(u) \, du$.

Consequently, the marginal hazard ratio is:

\begin{align}
\label{eq:analytic hazard ratio}
HR_M(t) 
&= \frac{\lambda_M(t|X=1)}{\lambda_M(t|X=0)} \notag \\
&= \exp(\beta_{C}) \left( \frac{1 + \theta A_{0,C}(t)}{1 + \theta A_{0,C}(t) \exp(\beta_{C})} \right)
\end{align}

First, note from equation \eqref{eq:analytic hazard function} that the true marginal hazard function does not follow a Cox PH structure. Second, from equation \eqref{eq:analytic hazard ratio}, we see that if $\theta=0$, then $HR_M(t)=\exp(\beta_{C})$. Note also that in this specific case, if $\theta \neq 0$, then $\lim_{t \to +\infty} HR_M(t) = 1$, which means that $HR_M(t)$ given by \eqref{eq:analytic hazard ratio} does not converge to the true conditional hazard ratio $\exp(\beta_{C})$.

We compared the theoretical value of $HR_M(t)$ to the hazard ratio estimated from two Cox models without adjustment for $U$. The first is a semi-parametric Cox model with a constant effect for $X$, as defined in equation \eqref{eq:unadjusted cox model}. The second is a semi-parametric Cox model with a time-dependent effect for $X$ unadjusted for $U$, specified as follows:

\begin{equation}
\label{eq:time dep cox model}
\lambda_{unadj}(t \mid X) = \lambda_{0,unadj}(t) \exp\left(\beta_{unadj}(t) X\right)
\end{equation}

In this model, the coefficient $\beta_{unadj}(t)$ is modeled using a natural cubic spline expansion:
\begin{equation*}
\displaystyle
\beta_{unadj}(t)=\sum_{j=1}^{k+1} a_j B_j(t)
\end{equation*}

where $k$ is the number of interior knots and $B_j(\cdot)$ denotes the $j^{\text{th}}$ basis function of the natural cubic spline.

To carry out the comparison, we simulated 1,000 datasets, each of size $n = 1000$, with covariates generated independently as $X \sim \mathcal{B}er(0.5)$ and $U \sim$ log-gamma(1,1) ($\theta=1$).  The  time to event $T$ was generated according to the conditional PHM in equation \eqref{eq:true model}, assuming a Weibull baseline risk, and the censoring time was generated from a Weibull distribution, right truncated at $t = 12.60$. We set the coefficients to $\beta_{C} = -0.6$ and $\beta_U = 1$ throughout these simulations.

\begin{figure*}
 \centering
  \includegraphics[width=15.8cm]{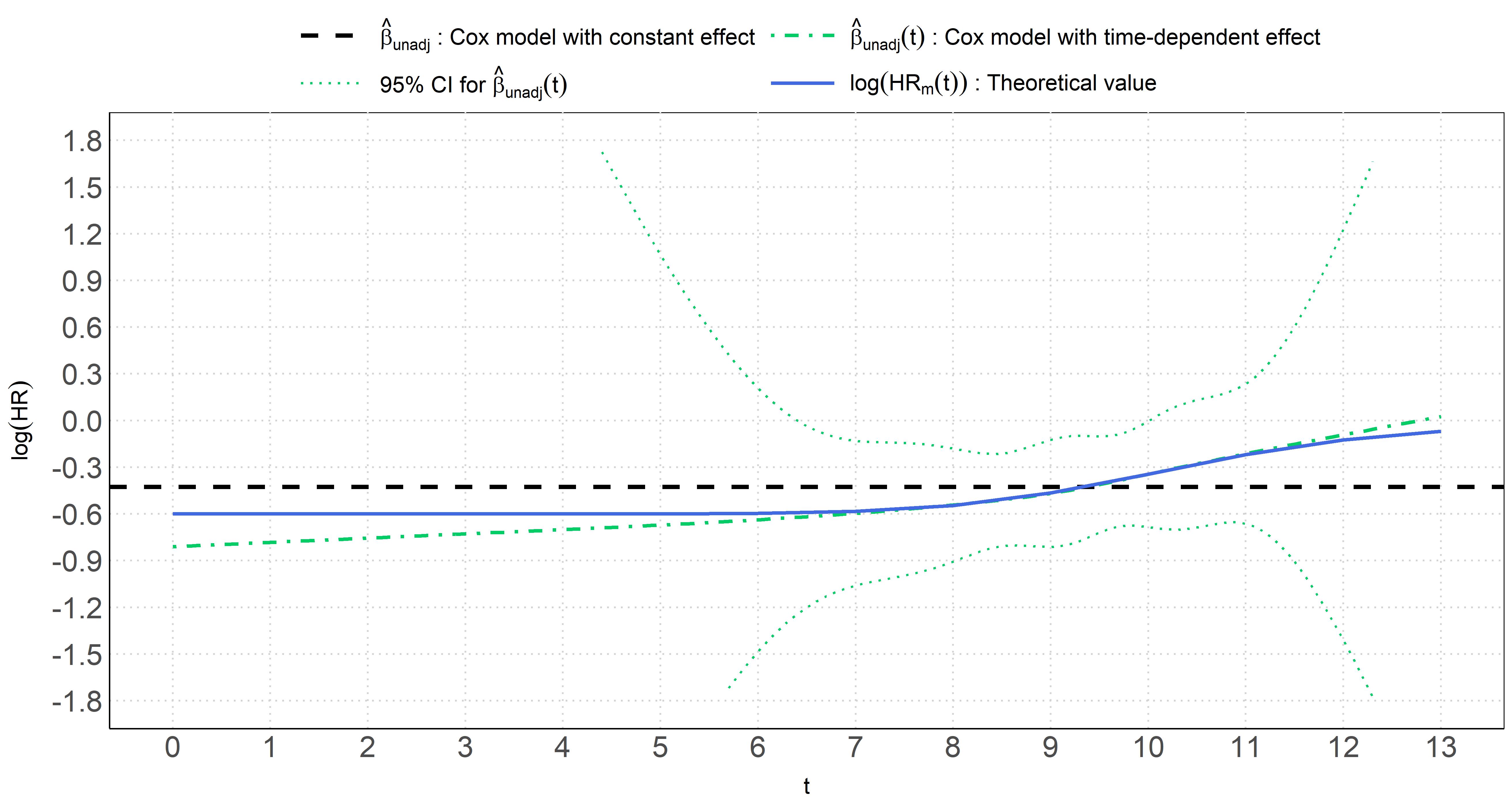}
  \centering
   \caption{Comparison of the true marginal log-hazard ratio (solid blue line) with estimates from an unadjusted Cox model with time-dependent effect (dotted-dashed green line, with 95\% CI in dotted green lines) and an unadjusted Cox model with constant effect (black dashed line). Estimates are based on 1,000 simulations under a gamma frailty model with $\beta_{C} = -0.6$ and $\beta_U = 1$.}
\label{fig3}
\end{figure*}

Figure~\ref{fig3} displays the theoretical marginal log-hazard ratio $\log(HR_M(t))$, along with estimates from both semi-parametric Cox models. As expected, the estimate $\hat{\beta}_{unadj}$ from the unadjusted model, which omits $U$, substantially overestimates the true conditional regression parameter $\beta_{C} = -0.6$. In contrast, the time-dependent estimate $\hat{\beta}_{unadj}(t)$ closely approximates the true marginal log-hazard ratio $\log(HR_M(t))$, despite the fact that the unadjusted Cox model with time-dependent effect is also misspecified, since the true marginal hazard function does not follow a Cox PH structure (equation~\eqref{eq:analytic hazard function}). The deviation between $\hat{\beta}_{unadj}(t)$ and $\log(HR_M(t))$ at early times can be attributed to the scarcity of events in this time window, with an average across the 1,000 replicates of only eight observed events before $t=6$, as reflected by the very large confidence interval.

\section{Properties of the unadjusted hazard ratio in a Cox PH model with constant effect}
\label{sec:Propertiesmarginal}

\subsection{Asymptotic properties}

This section presents a theoretical review of the properties of the partial likelihood estimator $\hat{\beta}_{unadj}$ obtained from the Cox model specified in equation \eqref{eq:unadjusted cox model}, which omits the covariate $U$ and assumes a constant effect of $X$ on the hazard function.

Let $T_{i}^{*}=\min(T_i,C_i)$ denote the observed time to event for a subject $i \in \{1,...,n\}$, where $T_i$ is the time to event governed by $\lambda_{C}(.)$ as defined in \eqref{eq:true model}, and $C_i$ is the censoring time assumed to be independent from $T_i$. Let $\delta_i=\mathds{1}(T_i\leq C_i)$ denote the failure indicator. Then the maximum partial likelihood estimator $\hat{\beta}_{unadj}$ is the value that maximizes the partial likelihood (Cox 1975)

\begin{equation}
\label{eq:partial-likelihood}
\displaystyle
L({\beta}_{X,M},X)=\prod_{i=1}^{n}\left[\frac{\exp(\beta_{unadj}X_i)}{\sum_{T_j^* \geq T_i^*}\exp(\beta_{unadj}X_j)}\right]^{\delta_i}
\end{equation}

Although the partial likelihood estimator $\hat{\beta}_{unadj}$ is well defined, the theoretical parameter $\beta_{unadj}$ in the model of equation \eqref{eq:unadjusted cox model} lacks a direct interpretation. This is because the true marginal log hazard ratio is, in general, a time-dependent function, and the true marginal hazards model does not typically follow a Cox PH structure.  

However, as an approximation, the partial likelihood estimate $\hat{\beta}_{unadj}$ can be viewed as the best-fitting parameter under the unadjusted model that omits $U$. 
Nonetheless, it can be shown \cite{struthers1986misspecified} that the partial likelihood estimator $\hat{\beta}_{unadj}$ is consistent for a well-defined constant, denoted by $\beta_{unadj}^*$, and that it is asymptotically normal. Importantly, we have the inequality \cite{daniel2021making}

\begin{equation}
\label{eq:shrinkagetoward0}
\left|\beta_{unadj}\right| \leq \left|\beta_{C}\right|
\end{equation}

This inequality highlights that, when a relevant covariate $U$ is omitted from the model, the unadjusted log-hazard ratio $\beta_{unadj}$ reflects an attenuated association between $X$ and the time-to-event outcome, and is generally smaller in magnitude than the true conditional effect $\beta_{C}$ that would be obtained if $U$ was included in the model.

Furthermore, the partial likelihood limit $\beta_{unadj}^*$ depends on the censoring distribution \cite{struthers1986misspecified} . While it is often suggested that $\exp(\beta_{unadj}^*)$ can be interpreted as a weighted average of the true, time-varying hazard ratio $HR_M(t)$ over the follow-up period, this interpretation does not hold in general \cite{vansteelandt2024assumption}. An important exception is when the unadjusted model includes only a single binary covariate \cite{stensrud2025use}, in this setting, $\exp(\beta_{unadj}^*)$ can be interpreted as a a weighted average of $HR_M(t)$ with weights depending on time and censoring rate.

\subsection{Hypothesis testing}

Consider the hypothesis test

\begin{equation*}
\label{eq:hypotheses}
\displaystyle
H_0: \beta_{C} = 0 \quad \text{vs} \quad H_1: \beta_{C} \neq 0
\end{equation*}

Under the null hypothesis $H_0$, it can be shown that the Wald test based on the partial likelihood estimator $\hat{\beta}_{unadj}$, obtained from the model that omits the covariate $U$, remains valid. Intuitively, this is because when $\beta_{C} = 0$, the covariate $X$ has no effect on the event time. As a result, survival up to time $t$ does not act as a collider in the causal diagram (see Figure~\ref{fig1}), and hence, omitting $U$ does not induce bias in testing for the effect of $X$.

More formally, under $H_0$, the asymptotic distribution of the estimator $\hat{\beta}_{C}$ obtained from the correctly specified model \eqref{eq:true model} coincides with that of $\hat{\beta}_{unadj}$ \cite{struthers1986misspecified} . This implies that the test statistic based on $\hat{\beta}_{unadj}$ follows the correct null distribution.

However, when $\beta_{C} \neq 0$, meaning that $X$ truly influences the hazard, the misspecification introduced by omitting $U$ causes $\hat{\beta}_{unadj}$ to be a biased and attenuated estimate of $\beta_{C}$. Consequently, the Wald test based on $\hat{\beta}_{unadj}$ loses power. 

\section{Accelerated Failure Time Model}
\label{sec:AFT}

AFT models are often considered as an alternative to the Cox PHM \cite{aalen2015does}. Unlike hazard ratios, which are subject to selection bias due to conditioning on survival, AFT models produce effect estimates that are collapsible and offer a clearer causal interpretation.

\subsection{General Form of the AFT Model}

Rather than modeling the hazard function, AFT models describe directly how covariates affect the logarithm of survival time. The general form of an AFT model including covariates $X$ and $U$ is given by:

\begin{equation}
\label{eq:aft-general}
\displaystyle
Y = \log(T) = \gamma_0 + \gamma_1 X + \gamma_2 U + \sigma \varepsilon
\end{equation}

where $\varepsilon$ is a mean-zero random error term that is independent of $(X,U)$. As in standard linear regression, the choice of the error distribution $\varepsilon$ determines the specific parametric form of the AFT model. 

\subsection{Interpretation of AFT Regression Coefficients}

The regression coefficients in the AFT model have a direct interpretation on the time scale. Specifically, taking the conditional expectations of $Y$ given $X$ and $U$ in ~\eqref{eq:aft-general} , and using the assumption that $\varepsilon \indep (X,U)$ we obtain:

\begin{equation}
\label{eq:expectAFT4}
\displaystyle
\mathbb{E}(\log(T)\mid X=1,U) - \mathbb{E}(\log(T)\mid X=0,U)= \gamma_1
\end{equation}

Thus, $\gamma_1$, often referred to as the "acceleration factor", indicates by how much on average (on a log scale) the event times are accelerated or decelerated in the group with $X=1$ compared to the group with $X=0$, conditionally on $U$.

\subsection{Collapsibility of the effect measure in the AFT Model}

Taking the conditional expectation of both sides of Equation~\eqref{eq:aft-general} given $X$ yields:

\begin{equation*}
\mathbb{E}(\log(T) \mid X) = \gamma_0 + \gamma_1 X + \gamma_2 \mathbb{E}(U \mid X) + \sigma \mathbb{E}(\varepsilon \mid X)
\end{equation*}

Since $U \indep X$ and $\epsilon \indep X$ by assumption, we have $\mathbb{E}(U \mid X)=\mathbb{E}(U)$ and $\mathbb{E}(\varepsilon \mid X)=\mathbb{E}(\varepsilon)=0$. Therefore, the above expression simplifies to:
\begin{equation}
\label{eq:expectAFT5}
\mathbb{E}(\log(T) \mid X) = \gamma_0^* + \gamma_1 X
\end{equation}
where $\gamma_0^*=\gamma_0+\gamma_2 \mathbb{E}(U)$  is a constant that does not depend on $X$.

Hence, the marginal effect measure in equation~\eqref{eq:expectAFT5}, which is $\gamma_1$, is equal to the conditional effect measure in equation~\eqref{eq:expectAFT4}. Thus, the AFT regression coefficient $\gamma_1$ is collapsible, provided that $\epsilon$ is independent of $(X,U)$.

\subsection{Equivalence between the AFT model and the Weibull PHM}

The Weibull distribution (with the exponential distribution as a special case) is unique among standard survival distributions in that it can be formulated both as a PHM and as an AFT model. In other words, a parametric PHM with a Weibull baseline hazard has an equivalent representation in the AFT framework, something that does not hold for other standard survival distributions.

Consider the model in Equation~\eqref{eq:true model}, and assume that the baseline hazard $\lambda_{0,C}(t)$ follows a Weibull distribution with scale and shape parameters denoted by $a$ and $b$, respectively, under the following parametrization:
\begin{equation*}
\lambda_{0,C}(t)=\frac{b}{a^b}t^{b-1}
\end{equation*}

Then

\begin{equation*}
S_{\text{C}}(t \mid X,U)
= \exp\!\left[ - \left(\frac{t}{a}\right)^{b} \exp(\beta_{\text{C}} X + \beta_U U) \right]
= \exp\!\left[ - \exp( -b\log a + \beta_{\text{C}} X + \beta_U U + b\log t) \right]
\end{equation*}

Therefore

\begin{align*}
\mathbb{P}\!\left( \log T \le t \,\middle|\, X,U \right)
&= \mathbb{P}\!\left( T \le e^{t} \,\middle|\, X,U \right) \\
&= 1 - \exp\!\left\{ - \exp( -b\log a + \beta_{\text{C}} X + \beta_U U + bt) \right\} \\
&= \mathbb{P}\!\left( W \le -b\log a + \beta_{\text{C}} X + \beta_U U + bt \,\middle|\, X,U \right) \\
&= \mathbb{P}\!\left( \log a - \frac{\beta_{\text{C}}}{b} X - \frac{\beta_U}{b} U + \frac{W}{b} \le t \,\middle|\, X,U \right)
\end{align*}

where $W$ follows a \textbf{standard extreme value distribution}, with CDF given by

\begin{equation*}
    F_W(w)=1-\exp\!\left\{-\exp(w)\right\}
\end{equation*}

This matches the structure of the AFT model given in Equation~\eqref{eq:aft-general}, with the following coefficients and error term: \begin{itemize} \item $\gamma_0 = \log(a)$ \item $\gamma_1 = -\dfrac{\beta_{C}}{b}$ \item $\gamma_2 = -\dfrac{\beta_{U}}{b}$ \item $\varepsilon = W$ \item $\sigma = \dfrac{1}{b}$ \end{itemize}

This demonstrates that a PHM with a Weibull baseline hazard is equivalent to an AFT model in which the error term follows a scaled extreme value distribution, specifically $\dfrac{\varepsilon}{b}$ where $\varepsilon$ follows a standard extreme value distribution.

\subsection{Robustness of the Regression Estimator in the Unadjusted AFT Model}

Let us assume that model \eqref{eq:aft-general} holds. We aim to estimate the unadjusted AFT model:

\begin{equation}
\label{Unadj-AFT}
Y = \log(T) = \gamma_0 + \gamma_1 X + W
\end{equation}

where the error term is denoted by $W$. Although the AFT regression coefficient $\gamma_1$ is collapsible, correctly estimating the unadjusted model \eqref{Unadj-AFT} requires a correct specification of the distribution of $W$. In practice, this is not feasible since the true error term is given by $\gamma_2 U + \sigma \varepsilon$ and $U$ is unobserved. However, several robust approaches have been proposed in the literature to address this issue.

One approach is to use symmetric error distributions, which tend to yield more robust estimators than asymmetric ones \cite{gould1988consistency}. In particular, the log-logistic distribution was reported to produce more robust estimates than the model with log-normal errors. The second approach consist of a flexible parametric AFT model. Specifically, model \eqref{Unadj-AFT} can be rewritten using the cumulative hazard functions as:

\begin{equation}
\label{Unadj-AFT4}
H_T(t \mid X) = H_{W^*}\left(t \exp(-\gamma_1 X)\right)
\end{equation}

where $W^*=\exp(\gamma_0+W)$, with $H_T$ and $H_{W^*}$ denoting the cumulative hazard functions of $T$ and $W^*$, respectively. If we assume that $W^*$ follows a Weibull distribution, with scale and shape parameters denoted by $a$ and $b$, respectively, then the log cumulative hazard function satisfies:

\begin{align}
\label{Unadj-AFT5}
\log(H_T(t \mid X))
&=\log\left(H_{W^*}(t \exp(-\gamma_1 X))\right) \notag \\
&= b \left[ \log(t \exp(-\gamma_1 X)) - \log(a) \right]
\end{align}

Therefore, in the Weibull model, the log cumulative hazard is a linear function of $\log(t \exp(-\gamma_1 X))$. Crowther et al.\cite{crowther2023flexible} propose a flexible parametric approach that generalizes the Weibull model by replacing the linear form with a smooth spline function $s\left(\log(t \exp(-\gamma_1 X))\right)$, thereby allowing for substantial flexibility. Thus, the flexible parametric AFT model can be written as:

\begin{align}
\label{Unadj-AFTSPLINES}
\log(H_T(t \mid X))=
s\left(\log(t \exp(-\gamma_1 X))\right)
\end{align}

\section{Frailty models for approximating the true conditional hazard ratio $\exp(\beta_{C})$}
\label{sec:FRAILTY}

Frailty models \cite{vaupel1979impact} are a natural extension of the Cox PHM that incorporate subject-specific random effects, called "frailties", to account for unobserved heterogeneity, typically arising from omitted covariates. The frailty term is represented by a random effect which, like explanatory covariates, influences the hazard multiplicatively. The frailty model takes the form:

\begin{align}
\lambda(t_i \mid \omega_i, X_i) 
&= \lambda_0(t_i) \omega_i \exp(\beta_{X,\omega} X_i) \notag \\
&= \lambda_0(t_i) \exp(\beta_{X,\omega} X_i + \log(\omega_i)) \label{eq:frailtymodel}
\end{align}

where $\omega_i$ denotes the individual-specific frailty.

Frailty models are used in the hope that the random effect $\omega_i$ can help correct the bias caused by omitting an unobserved variable $U$. The idea is that $\omega_i$ (or more precisely $\log(\omega_i)$) may capture the variability induced by the omitted term $\beta_U U_i$. For instance, suppose the unobserved covariate follows $U_i \sim \mathcal{N}(0,1)$, so that the omitted component in the linear predictor, $\beta_U U_i$, follows a normal distribution $\mathcal{N}(0, \beta_{U_i}^2)$. If we fit a frailty model assuming a normally distributed log-frailty, i.e., $\log(\omega_i) \sim \mathcal{N}(0, \sigma_{\omega}^2)$, then the variance $\sigma_{\omega}^2$ should absorb the variability induced by $\beta_U U_i$, and the coefficient estimate $\hat{\beta}_{X,\omega}$ will approximate the true conditional effect $\beta_{C}$. Parametric frailty models are estimated by maximizing the marginal likelihood computed by integrating the conditional likelihood over the frailty distribution \cite{rondeau2003maximum}. For semi-parametric frailty models, a penalized partial likelihood estimation approach has been proposed \cite{therneau2000cox}. As the marginal likelihood has no closed form when the frailty is log normal, the gamma distribution is preferred for the frailty, leading to a closed form for the marginal model as shown in formula \eqref{eq:analytic hazard function}. 
While this distribution may not match the true distribution of $\beta_U U_i$, we expect that the frailty term compensates for omitted covariate, and that $\hat{\beta}_{X,\omega}$ still provides a reasonable approximation to the true effect $\beta_{C}$.

\section{Survival difference between groups}
\label{sec:Survivaldifferences}

Survival differences between treatment groups are proposed as an alternative measure of association\cite{stensrud2020test}, as survival probabilities reflect absolute risks over time, rather than instantaneous risks conditional on being at risk at a particular time point. Unlike hazard ratios, the survival difference is collapsible.

The marginal survival difference, for all $ t > 0 $, can be expressed as:

\begin{align}
    S_M(t \mid X=1) - S_M(t \mid X=0) 
&= \int \left[ S_{C}(t \mid X=1, U) - S_{C}(t \mid X=0, U) \right] f_U(u) \, du \nonumber \\
&= \mathbb{E}_U \left( S_{C}(t \mid X=1, U) - S_{C}(t \mid X=0, U) \right) \label{eq:survivalmarginaldifference}
\end{align}

This expression highlights that survival differences are a \emph{collapsible} measure of association: the marginal difference corresponds to the average of the conditional differences over the distribution of $U$. However, survival differences are not \emph{strictly collapsible}, since the conditional measure depends on $U$. In contrast, this collapsibility property does not hold for the survival ratio.

Note that the marginal survival function estimated from the unadjusted Cox model defined in \eqref{eq:unadjusted cox model}:
\begin{equation}
\label{eq:unadjustedmarginalsurvival}
\hat{S}_{unadj}(t \mid X=x) = \exp\left(-\hat{A}_{0,unadj}(t)\exp(\hat{\beta}_{unadj} x)\right)
\end{equation}

is a biased estimator of $S_M(t \mid X=x)$ since we have previously shown that this model is misspecified.

When $ X $ is binary, the marginal survival functions can be estimated nonparametrically using the Kaplan-Meier estimator, and their difference directly reflects the contrast between groups $ X = 1 $ and $ X = 0 $. However, when $ X $ is continuous, estimation requires modeling the survival function over a range of covariate values. In such cases, an unadjusted Cox model with a time-dependent effect for $X$, as defined in \eqref{eq:time dep cox model}, can be employed, since we have shown in Figure  \ref{fig3} that it can approximate the true marginal hazard. This estimator is given by\cite{thomas2014tutorial}

\begin{equation}
\label{eq:timedepcoxdmarginalsurvival}
\hat{S}_M(t \mid X=x) = \exp\left(-\sum_{t_i \leq t}{\frac{\delta_i\exp(\hat{\beta}_{unadj}(t_i)x)}{\sum_{j=1}^n \mathds{1}(T_j \geq T_i)\exp(\hat{\beta}_{unadj}(t_i)x_j)}}\right)
\end{equation}

This approach requires selecting a finite and clinically relevant set of $ X $ values to compare, along with a set of time points at which the survival differences will be computed. 
For the Kaplan-Meier estimator, asymptotic standard deviations can be computed using Greenwood's formula. The variance of the survival differences is the sum of the variances since the two groups $X=1$ and $X=0$ are not overlapping. For the Cox-based estimator, asymptotic standard deviations can be obtained via bootstrap. For each bootstrap sample, the unadjusted Cox model with time-dependent effect of the exposure is re-estimated, and the survival difference is computed using the formula~\eqref{eq:timedepcoxdmarginalsurvival}.

\section{Simulations}
\label{sec:SIMU}

We conducted simulations to quantify the non-collapsibility of parametric (Weibull) and semi-parametric (Cox) PHM, and to evaluate the robustness of the gamma frailty model under different distributions of the omitted covariate $U$. We further examined the robustness of the unadjusted AFT model to the omission of $U$ under various assumptions regarding the distribution of the error term. Finally, we investigated whether marginal survival differences, when estimated using the Kaplan-Meier estimator or an unadjusted Cox model with time-dependent effect of the exposure, adequately approximate the average conditional survival difference over the distribution of $U$.

\subsection{Design}

For each scenario, we simulated 1000 datasets, each consisting of $n = 1000$ individuals. The covariates were independently generated as $X \sim \mathcal{B}er(0.5)$ and $U  \sim \mathcal{N}(0,1)$. Complementary simulations were performed with $U \sim$ log-gamma$(1,1)$  and $U\sim\mathcal{B}er(0.5)$ to evaluate the impact of the distribution of $U$ on the bias of the na\"{\i}ve unadjusted models and the robustness of the gamma frailty models. The observed time-to-event is defined as $T^* = \min(T, C)$, where $C$ is the censoring time, generated from a Weibull
distribution, right truncated at $t = 12.60$, corresponding to the maximum follow-up time observed in RTOG9202\cite{lawton2017duration}. The true event time $T$ was generated according to the conditional PHM in equation \eqref{eq:true model}, using the cumulative hazard inversion method \cite{brilleman2021simulating}. The regression coefficients were set to $\beta_{C} = -0.6$ and $\beta_U \in \{0.2,0.4,0.8,1\}$, with the aim of examining the impact of varying the intensity of the association between the omitted covariate $U$ and the event risk.

The baseline hazard function $\lambda_{0,C}(t)$ was assumed to follow a Weibull distribution with scale and shape parameters $a$ and $b$, respectively. These parameters were chosen for each scenario so that the resulting censoring proportion matched that observed in the RTOG9202 study, approximately $50\%$.

\subsection{Estimated models}

First, we estimated models that do not correct for the omission of the covariate $U$. These include two PH models adjusted only for $X$, specified as

\begin{equation*}
\label{eq:unadjusted cox model SIMU}
\lambda_{unadj}(T_i \mid X_i) = \lambda_{0,unadj}(T_i)\exp(\beta_{unadj}X_i)
\end{equation*}

The first model is a semi-parametric Cox model estimated via partial likelihood, using the \textbf{coxph} function from the \textbf{survival} package in R\cite{therneau2024package}. The second model is a parametric PH model in which $\lambda_{0,unadj}(t)$ is assumed to follow a Weibull distribution, for which estimation was performed using the \textbf{phreg} function from the \textbf{eha} package in R\cite{brostrom2024eha}.

We also estimated the following the unadjusted AFT model:

\begin{equation*}
Y_i = \log(T_i) = \gamma_0 + \gamma_1 X_i + \sigma \varepsilon_i
\end{equation*}

Four versions of this model were estimated, varying the distributional assumption for the error term $\varepsilon_i$. In the first specification, $\varepsilon_i$ follows a scaled standard extreme value distribution. In the second, $\varepsilon_i \sim \mathcal{N}(0,1)$. The third one assumes that $\varepsilon_i$ follows a log-logistic distribution. These three models were estimated using the \textbf{aftreg} function from the \textbf{eha} package. The fourth model is a flexible parametric AFT model in which the cumulative hazard function is modeled using restricted cubic splines. This model was estimated using the \textbf{aft} function from the \textbf{rstpm2} package in R\cite{clements2025rstpm2}.

We then estimated PH models with a subject-specific frailty: 

\begin{equation*}
\lambda(t_i \mid \omega_i, X_i) = \lambda_0(t_i) \omega_i \exp(\beta_{X,\omega} X_i)
\end{equation*}

where $\omega_i$ follows a Gamma distribution with parameters $(1/\theta, 1/\theta)$. Two variants of this model were estimated: a semi-parametric frailty model estimated via penalized partial likelihood (using the \textbf{coxph} function from the \textbf{survival} package) and a parametric frailty model assuming a Weibull baseline hazard, estimated (using the \textbf{frailtyPenal} function from the \textbf{frailtypack} package in R\cite{rondeau2025frailtypack}).  For all these methods, the confidence interval for the estimated regression coefficient of $X$ was computed using the asymptotic standard error provided by the software, based on the inverse Hessian.\\

The marginal survival differences, $S_M(t \mid X=1) - S_M(t \mid X=0)$, were estimated using two methods: a nonparametric approach using the Kaplan-Meier estimator (implemented via the \textbf{survfit} function from the \textbf{survival} package), and using a Cox model with a time-dependent effect for $X$ unadjusted for $U$, defined by ~\eqref{eq:time dep cox model}, with the survival differences computed using the formula ~\eqref{eq:timedepcoxdmarginalsurvival}. These differences were evaluated at three time points, $t = 8$, $9$, and $10$, corresponding respectively to the 25th, 50th, and 75th percentiles of the observed event times. For the Kaplan-Meier-based estimator, asymptotic standard deviations were computed using the Greenwood's formula. For the estimator based on the unadjusted Cox model with a time-dependent effect for $X$, asymptotic standard deviations were obtained via nonparametric bootstrap using B $=500$ resampled datasets. Due to this computational cost, the evaluation of survival differences was performed using only $200$ datasets with $n=300$ individuals and $U\sim \mathcal{N}(0,1)$.

\subsection{Results}
 
\begin{center}
\begin{figure*}
 \centering
  \includegraphics[width=15.8cm]{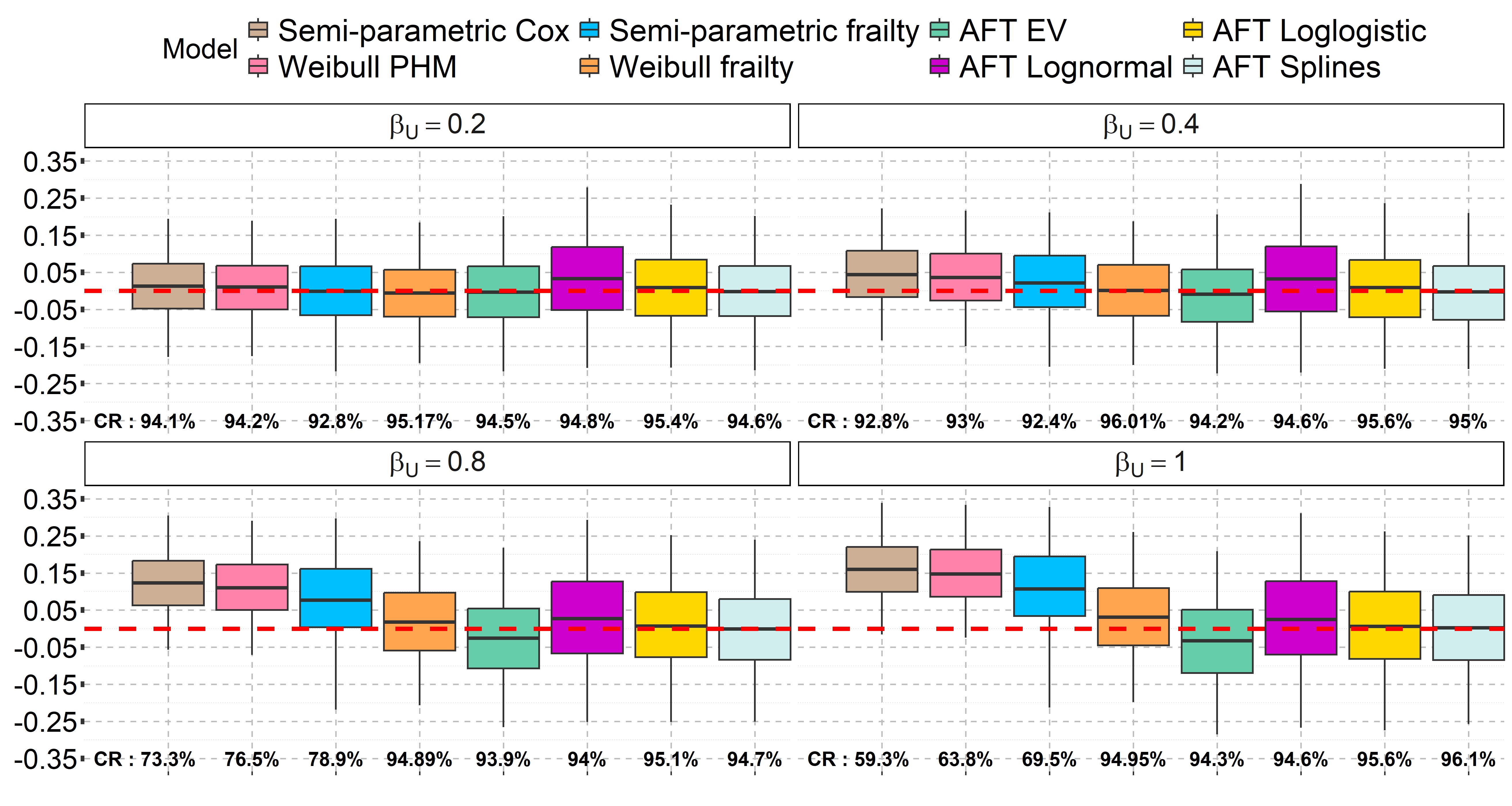}
  \centering
 \caption{Results of the simulation study with $U \sim \mathcal{N}(0,1)$. For each $\beta_U \in \{0.2, 0.4, 0.8, 1\}$, boxplots display the 2.5\%, 25\%, 75\%, and 97.5\% quantiles, as well as the mean, of the difference between the true and estimated values of the regression coefficient from the following models, ordered from left to right as: the unadjusted semi-parametric Cox model, the unadjusted Weibull PHM, the semi-parametric frailty model, the Weibull frailty model, the unadjusted AFT model with a standard extreme value error distribution (AFT EV), the unadjusted AFT model with a standard Gaussian error distribution (AFT Lognormal), the unadjusted AFT model with a log-logistic error distribution (AFT Loglogistic), and the flexible parametric AFT model based on splines (AFT Splines). To facilitate visual comparison, the AFT estimates were rescaled by a factor of 10 to match the scale of the other models. The coverage rates (CR) of the $95\%$ confidence intervals are given at the bottom of each boxplot. The dashed red horizontal line represents the line $y = 0$. All results are based on 1,000 simulated replicates.}
\label{fig4}
\end{figure*}
\end{center}

\begin{figure*}
 \centering
  \includegraphics[width=15.8cm]{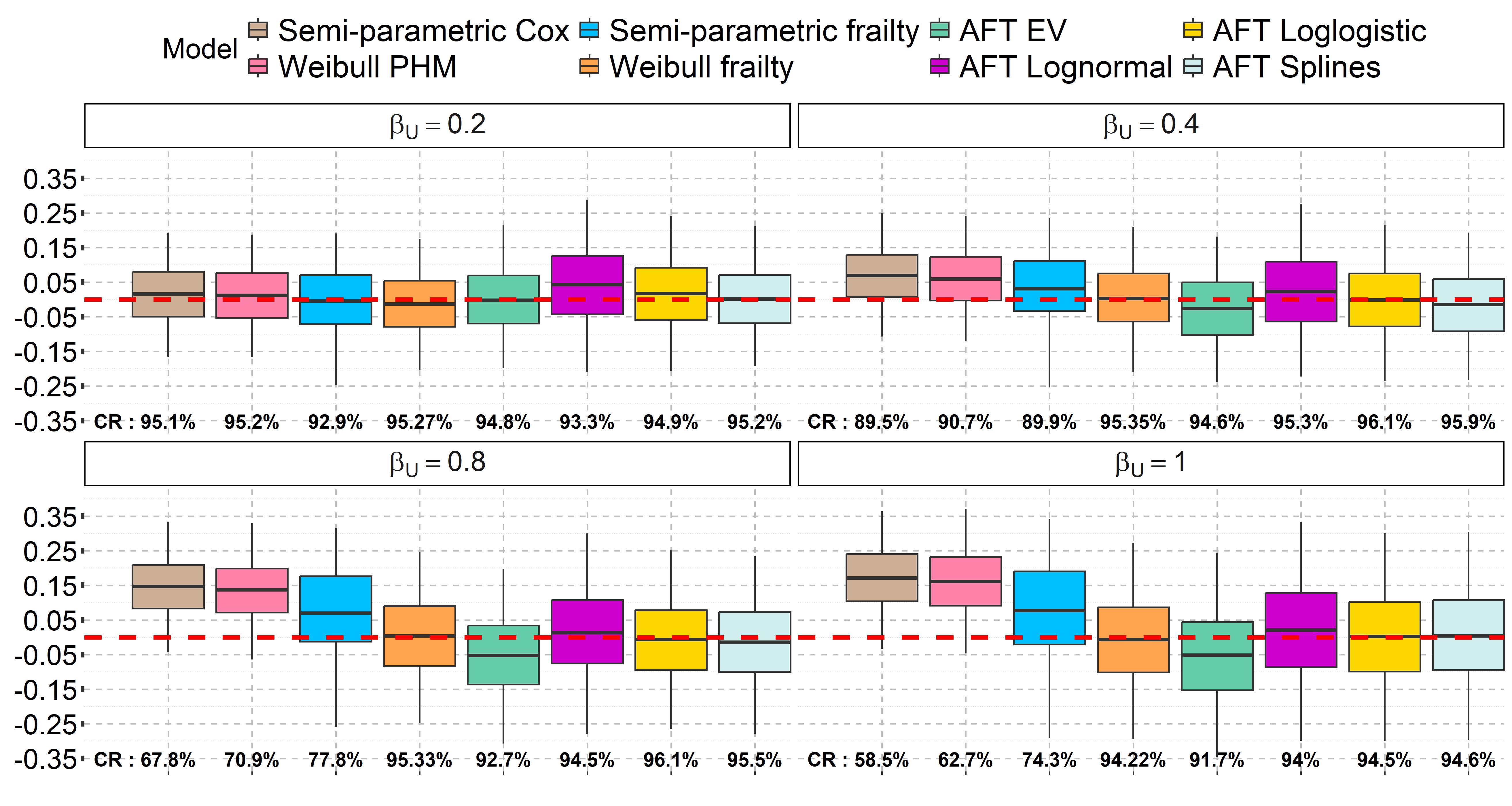}
  \centering
  \caption{Results of the simulation study with $U$ following a log-gamma$(1,1)$ distribution. For each $\beta_U \in \{0.2, 0.4, 0.8, 1\}$, boxplots display the 2.5\%, 25\%, 75\%, and 97.5\% quantiles, as well as the mean, of the difference between the true and estimated values of the regression coefficient from the following models, ordered from left to right as: the unadjusted semi-parametric Cox model, the unadjusted Weibull PHM, the semi-parametric frailty model, the Weibull frailty model, the unadjusted AFT model with a standard extreme value error distribution (AFT EV), the unadjusted AFT model with a standard Gaussian error distribution (AFT Lognormal), the unadjusted AFT model with a log-logistic error distribution (AFT Loglogistic), and the flexible parametric AFT model based on splines (AFT Splines). To facilitate visual comparison, the AFT estimates were rescaled by a factor of 10 to match the scale of the other models. The coverage rates (CR) of the $95\%$ confidence intervals are given at the bottom of each boxplot. The dashed red horizontal line represents the line $y = 0$. All results are based on 1,000 simulated replicates.}
\label{fig5}
\end{figure*}

\begin{figure*}
 \centering
  \includegraphics[width=15.8cm]{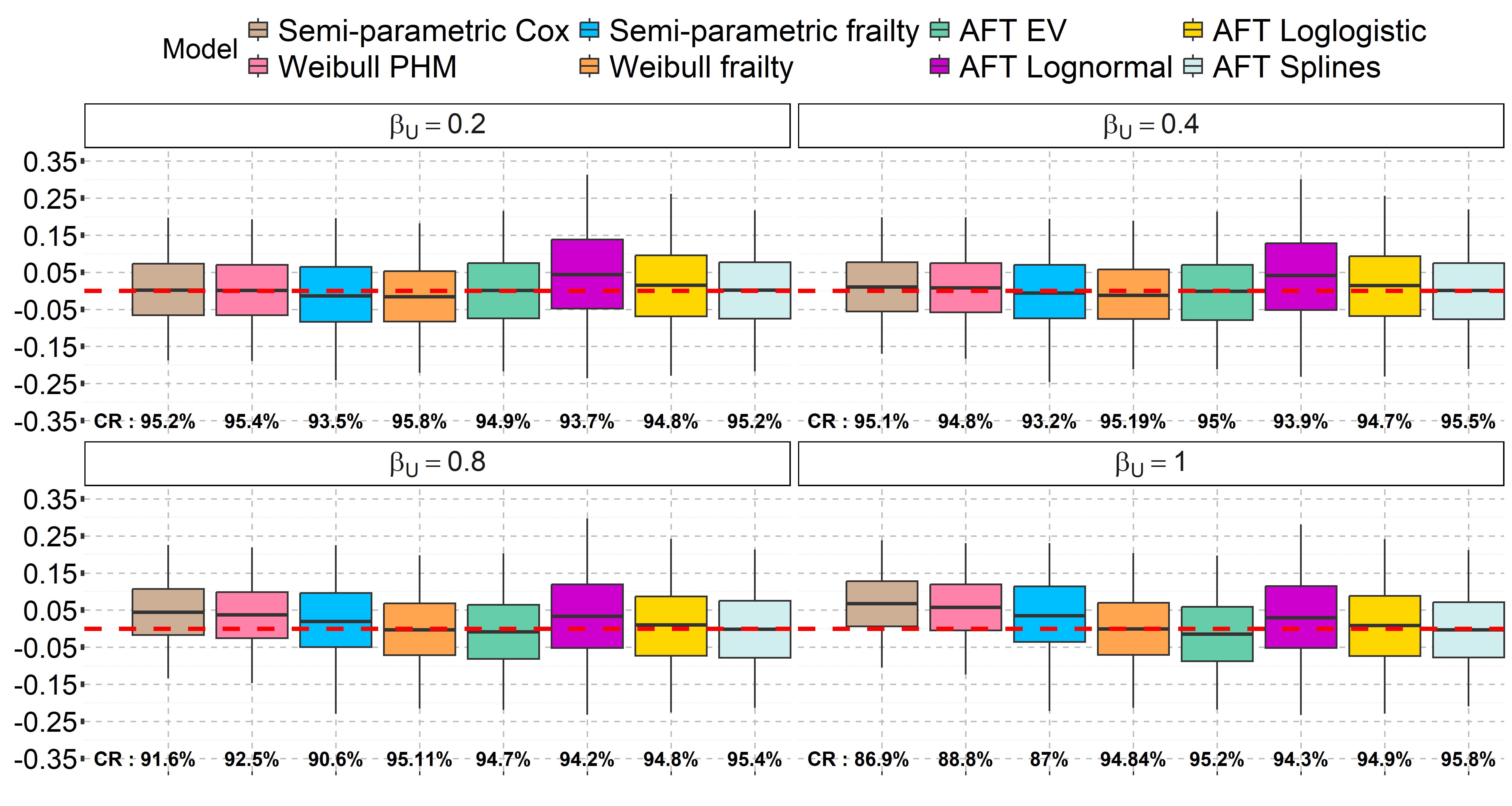}
  \centering
  \caption{Results of the simulation study with $U \sim \mathcal{B}er(0.5)$. For each $\beta_U \in \{0.2, 0.4, 0.8, 1\}$, boxplots display the 2.5\%, 25\%, 75\%, and 97.5\% quantiles, as well as the mean, of the difference between the true and estimated values of the regression coefficient from the following models, ordered from left to right as: the unadjusted semi-parametric Cox model, the unadjusted Weibull PHM, the semi-parametric frailty model, the Weibull frailty model, the unadjusted AFT model with a standard extreme value error distribution (AFT EV), the unadjusted AFT model with a standard Gaussian error distribution (AFT Lognormal), the unadjusted AFT model with a log-logistic error distribution (AFT Loglogistic), and the flexible parametric AFT model based on splines (AFT Splines). To facilitate visual comparison, the AFT estimates were rescaled by a factor of 10 to match the scale of the other models. The coverage rates (CR) of the $95\%$ confidence intervals are given at the bottom of each boxplot. The dashed red horizontal line represents the line $y = 0$. All results are based on 1,000 simulated replicates.}
\label{fig6}
\end{figure*}

Figures~\ref{fig4}, ~\ref{fig5} and ~\ref{fig6} present the results of the simulation study for the unadjusted semi-parametric Cox model, the unadjusted PHM, the frailty models, and the unadjusted AFT model with different distributions of the error, where the omitted covariate  $U \sim \mathcal{N}(0,1)$ in Figure \ref{fig4}, $U \sim \text{log-gamma(1,1)}$ in Figure \ref{fig5} and $U \sim \mathcal{B}er(0.5)$ in Figure \ref{fig6}. In these figures, we systematically evaluate all the  models across various scenarios, corresponding to increasing values of the coefficient $\beta_U$ (i.e., $0.2$, $0.4$, $0.8$, and $1$). Detailed numerical results are given on Tables $S1,S2$ and $S3$ in the supplementary material.

As shown in Figure \ref{fig4} where $U \sim \mathcal{N}(0,1)$, the unadjusted semi-parametric Cox model and the unadjusted Weibull PHM, tend to produce more biased estimates and exhibit poor coverage rates as the effect of the omitted covariate increases. For instance, when $\beta_U = 0.2$ (i.e., when the effect of the unobserved $U$ is weak), the unadjusted semi-parametric Cox model yields a standardized bias of $14\%$ (Table S1), with a coverage rate of $94.1\%$. However, when $\beta_U = 1$, the standardized bias increases to $178\%$ (Table S1), and the coverage rate drops to $59.3\%$.

The unadjusted AFT model with standard extreme value distribution (AFT EV) in Figure \ref{fig4} where $U \sim \mathcal{N}(0,1)$ provides relatively robust estimates, with only minor bias and generally satisfactory coverage rates. However, in Figure \ref{fig5} where $U$ follows a log-Gamma distribution, the coverage rate for this model drops to $91.7\%$ when $\beta_U = 1$. This can be explained by misspecification of the error distribution when $U$ is omitted. On the other hand, the unadjusted AFT model with log-normal (AFT Lognormal) or log-logistic distribution (AFT Loglogistic), and the flexible parametric AFT model (AFT Splines), all yields robust results for all values of $\beta_U$ and across all the three scenarios. 

Frailty models in Figure \ref{fig4} where $U \sim \mathcal{N}(0,1)$ generally perform better than the uncorrected models. However, the results for the semi-parametric frailty model are not entirely satisfactory. Notably, the asymptotic standard deviation is consistently underestimated compared to the empirical standard deviation (Table S1), which affects the coverage rate.

The parametric frailty model, where the baseline hazard function has been correctly specified, performs relatively well and shows robust results. For example, when $U \sim \mathcal{N}(0,1)$ and $\beta_U = 1$ (Figure \ref{fig4} and Table S1), the standardized bias is limited to $26\%$, with a nominal coverage rate. Nevertheless, some bias persists, which can be attributed to a misspecification of the frailty distribution: the true distribution of $U$ is Gaussian, while the model assumes a Gamma-distributed frailty. Figure \ref{fig5} supports this interpretation. Indeed, when the omitted covariate $U$ follows a log-Gamma distribution (i.e., when the assumed frailty distribution aligns with the true distribution), the parametric frailty model demonstrates robust performance across all values of $\beta_U$, with a standardized bias of $6\%$ when $\beta_U=1$ (Table S2).

Figure \ref{fig6} shows the results when the omitted covariate follows a Bernoulli distribution: $U \sim \mathcal{B}er(0.5)$. The omission of a binary covariate leads to smaller biases for the naïve unadjusted  models compared to the omission of a continuous covariate (Figures \ref{fig4} and \ref{fig5}) and the Weibull frailty model with gamma frailty appears robust to this misspecication of the frailty. The four AFT models also have good behavior in this scenario.

\begin{figure*}
 \centering
  \includegraphics[width=15.8cm]{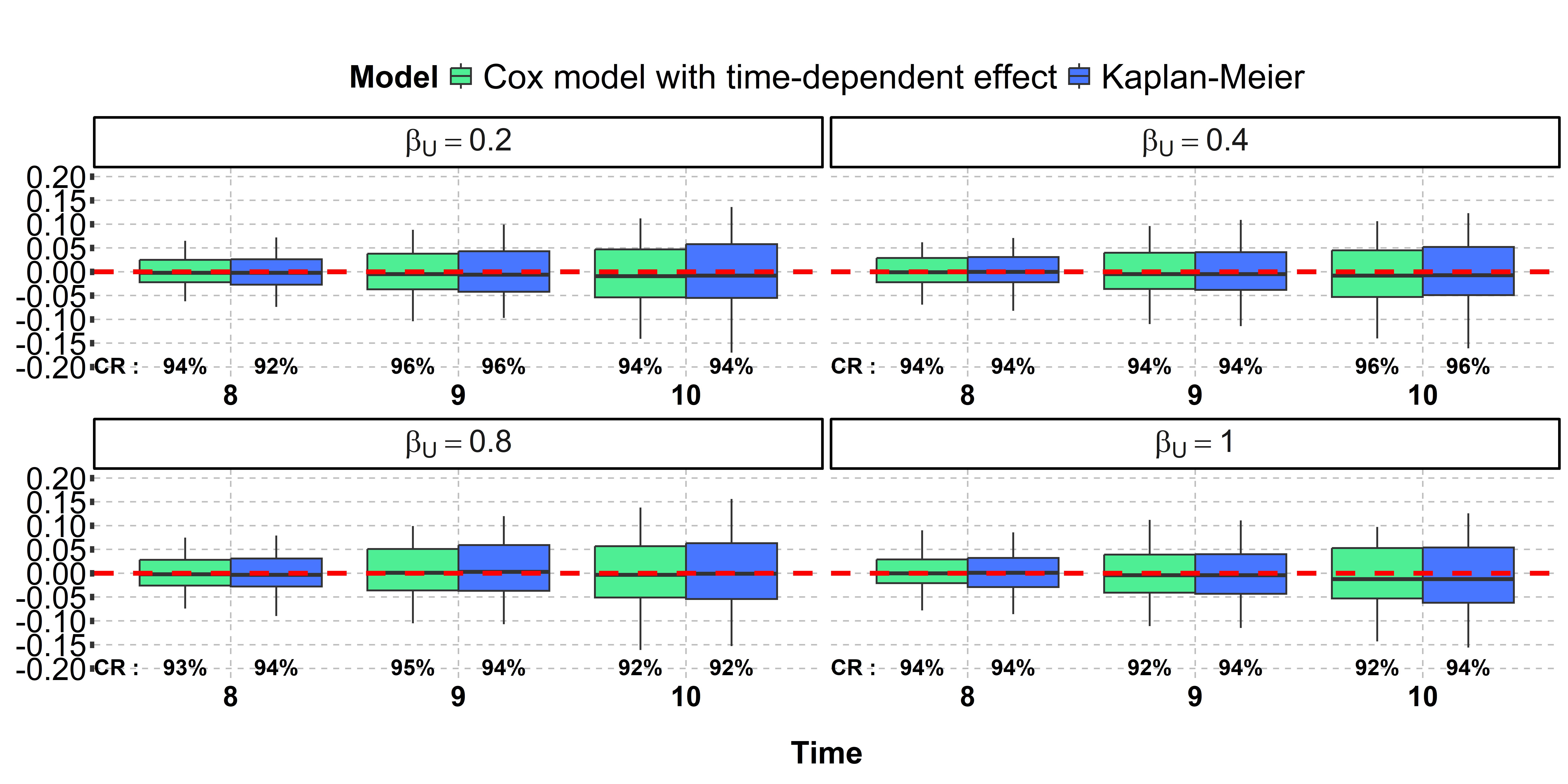}
  \centering
  \caption{Results of the simulation study with $U \sim \mathcal{N}(0,1)$. 
For each $\beta_U \in \{0.2, 0.4, 0.8, 1\}$, boxplots display the 2.5\%, 25\%, 75\%, and 97.5\% quantiles, together with the mean, of the difference between the estimated and true survival difference between exposure groups, obtained using the Kaplan-Meier estimator (right boxplot) and a Cox model with time-dependent exposure effect (left boxplot), for the three time points $t = 8, 9,$ and $10$. The coverage rates (CR) of the $95\%$ confidence intervals are given at the bottom of each boxplot. The dashed red horizontal line indicates $y = 0$. Results are based on 200 simulated replicates.}
\label{fig7}
\end{figure*}

Figure~\ref{fig7} displays the marginal survival differences, estimated using either the Kaplan-Meier estimator or a Cox model with a time-dependent effect of the exposure, unadjusted for $U$, for the three time points $t = 8, 9,$ and $10$. These estimators show little to no bias and maintain good coverage rates for all values of $\beta_U$. The true values of the marginal survival differences, given in~\eqref{eq:survivalmarginaldifference}, were approximated via numerical integration.

\section{Application}

We applied the models evaluated in the simulation study to the \textbf{R}adiation \textbf{T}herapy \textbf{O}ncology \textbf{G}roup 92-02 (RTOG 92-02) dataset in order to assess the performance of the proposed alternative approaches relative to the uncorrected models in a real-world setting.

\subsection{Data}

The Radiation Therapy Oncology Group (RTOG) Protocol 92-02 was a randomized trial evaluating the effect of long-term adjuvant androgen deprivation (AD) following initial AD with external-beam radiotherapy (RT) in patients with locally advanced prostate cancer and prostate-specific antigen (PSA) levels below 150 ng/mL. All patients received a total of 4 months of goserelin and flutamide, administered 2 months before and 2 months during RT. Radiotherapy involved a dose of 65–70 Gy to the prostate and 44–50 Gy to the pelvic lymph nodes. Patients were randomly assigned to one of two arms: no additional therapy, or 24 additional months of goserelin. The dataset includes $n = 1116$ patients. Subjects were followed every 3 months during the first year, every 4 months during the second year, every 6 months during years 3 to 5, and annually thereafter. The time-to-event outcome is defined as the number of years from the end of radiotherapy to either local or distant recurrence, or death. The proportion of observed events was $51.35\%$ ($N=573$). The main exposure covariate, denoted $X$, is binary: $X = 0$ indicates no additional therapy ($N=554$), and $X = 1$ indicates 24 months of extended hormonotherapy ($N=562$). The adjustment variables were four baseline covariates measured before treatment:

\begin{itemize}
     \item \textbf{Gleason score} with three levels: $<7$, $=7$, and $>7$, with frequencies of 457, 386, and 273, respectively.
    \item \textbf{Tumour stage} with two levels: T2 ($N = 519$), when the tumour is confined within the prostate, and T3-T4 ($N = 597$), when the tumour extends through the prostatic capsule or invades adjacent structures other than the seminal vesicles.
    \item \textbf{PSA:} pre-treatment prostate-specific antigen in ng/mL, transformed as $\log(\text{PSA} + 0.1)$.
   \item \textbf{Age at entry} (range:   $43$ to $88$ years, with a median of $70$ years) centered and divided by $10$.
\end{itemize}

\subsection{Estimated models}

We estimated the eight models previously examined in the simulation study. In the application, we distinguished two cases: the first where we adjusted only for the exposure variable $X$, and the second where we adjusted for both the exposure and the four additional baseline covariates described above (Gleason score, tumour stage, PSA, and age).

\subsection{Model checking}

Model checking was performed for all eight models, following the procedures described in Chapters 4 and 7 of Collett D.\cite{collett2023modelling}. The plots of the cumulative hazard function of the Cox-Snell residuals against their corresponding values were examined for each model, and all indicated a satisfactory overall fit. In addition, plots of martingale residuals versus the quantitative covariates PSA and age did not reveal any systematic patterns, indicating that no transformation of these covariates was necessary. Diagnostic plots for the adjusted semi-parametric Cox model are displayed in Figures S1, S2 and S3 of the Supplementary Material.

\subsection{Results}

Table~\ref{tab:appli-models} reports the models' estimates of the treatment effect on the RTOG9202 dataset with and without adjustement on baseline covariates.
All methods estimated a significant protective effect of the treatment, but the two naïve unadjusted models yield estimates that differ notably from those of the frailty models. For example, the estimate from the semi-parametric Cox model is $-0.173$, which is closer to zero than the estimate from the Weibull frailty model ($-0.233$), which performed well in simulations. As expected, the estimates from the Weibull frailty model remain relatively stable across both adjusted and unadjusted settings.

In contrast, the estimates from the semi-parametric Cox and Weibull PHM change considerably when adjusting for the baseline covariates and becomes much closer to the estimates from the Weibull frailty model. This suggests that the four baseline variables capture a substantial part of the heterogeneity previously absorbed by the frailty term as supported by the steep decline of the estimated variance of the frailty after adjustment (from $\hat{\theta}=0.469$ (SE = $0.461$) to $\hat{\theta}=0.04$ (SE = $0.331$)).

As expected from the simulation study, the AFT model with log-normal or log-logistic error distributions exhibits very stable results regardless of adjustment, while the estimate of treatment effect from the AFT model with extreme value error is more sensitive to the adjustment. For the flexible parametric AFT model (AFT Splines), the adjusted treatment effect differs slightly from the unadjusted one. 
Although this model showed robust performance in the simulation study, it may suffer from numerical instability, as varying the number of degrees of freedom of the spline function produced considerably different estimates of the treatment effect. In addition, using the age variable or its scaled version resulted in different estimates and different likelihood values.

Table~\ref{tab:appli-surv} presents the estimates of the marginal survival differences between treatment groups on the RTOG9202 dataset, estimated using either the Kaplan-Meier estimator or a Cox model with a time-dependent effect of the treatment, without adjustment on baseline covariates, for the three time points $t_{25} = 2.8, t_{50} = 5.3$ and $t_{75} = 7.8$ corresponding respectively to the 25th, 50th, and 75th percentiles of the observed event times.

\section{Discussion}

We have illustrated and quantified the built-in selection bias due to omitted covariate in the Cox model. To overcome this problem we have evaluated 3 approaches.
If the targeted measure of association is the hazard ratio and the impact of the missing covariates is expected to be low, the hazard ratio estimated by the Cox model is robust. When the missing covariates can have a large effect, parametric PHM with subject-specific gamma frailty is a useful alternative approach. We demonstrated that the models with subject-specific frailty are robust to misspecified frailty distribution (including when the true distribution is discrete). However, it requires to well characterize the baseline risk function because the semi-parametric frailty model exhibits poor results, due to the under-estimation of the standard error of the estimates and also a bias for the targeted parameter estimate when the impact of the missing covariate and thus the variance of the frailty increase. These poor results can be explained because these models include subject-specific frailty while frailty models were initially proposed to deal with clustered data. We found that parametric model with subject-specific frailty have good behavior but that the estimation of semi-parametric model with subject specific frailty is less stable. Additional simulations, not shown here, were conducted to evaluate a semi-parametric model with subject-specific frailty estimated via penalized likelihood (implemented in the \textbf{frailtyPenal} function of the \textbf{frailtypack} package in R \cite{rondeau2025frailtypack}). However, this model performed poorly for the same reason, namely that such models were originally designed to account for cluster-specific rather than subject-specific frailty. 

Otherwise, AFT models provide a collapsible and unique summary measure of the exposure effect. In both the simulation study and the real data application, the AFT models with log-normal or log-logistic errors appeared to be the most robust, regardless of the distribution of the unmeasured covariate. However, as these are parametric models, a careful evaluation of their underlying assumptions remains necessary (Chapter 7 of Collett D.\cite{collett2023modelling}).

Finally, as previously suggested\cite{stensrud2025use}, the difference in survival functions between exposure groups at selected time points can be estimated using either Kaplan-Meier estimates for binary exposures, or Cox model with spline-based time-dependent exposure effect for quantitative exposures or when the model includes adjustment covariates. We demonstrated the collapsibility of this association measure and the robustness of the estimates using the Cox model with spline-based time-dependent effect while it is not the true model. The main drawback of this approach is the selection of the time points and, possibly, of the values of adjustment variables.

This article provides an overview of key results from the literature, supported by analytical demonstrations of the main formulas. We also present a simulation-based evaluation quantifying the bias of the semi-parametric Cox and Weibull PH models and comparing their performance with several alternative approaches. Although the limitations of standard models and the potential robustness of alternative methods have been discussed previously, to our knowledge, no prior study has quantified these biases through simulations or assessed the extent to which alternative approaches can correct them. Furthermore, we demonstrated the practical usefulness of the proposed methods through an application using real data from the Radiation Therapy Oncology Group (RTOG 9202)\cite{lawton2017duration} randomized controlled trial.
For future work, it would be of interest to improve the inference procedure for the semi-parametric frailty model to better accommodate subject-specific frailty, which would relax the parametric assumption on the baseline hazard and enable greater flexibility in baseline distributions.

\clearpage

\vspace{4em}

\begin{table*}[h] 
\centering
\caption{Models' estimates of the treatment effect on the RTOG9202 dataset with and without adjustement on baseline covariates.} 
\label{tab:appli-models} 

\vspace{0.3cm}
\begingroup
\fontsize{8}{10}\selectfont

\resizebox{\ifdim\width>\linewidth\linewidth\else\width\fi}{!}{
\begin{tabular}{l r r r r r r}
\toprule
\textbf{Model} & \textbf{Unadjusted Estimate} & \textbf{SE} & 
\textbf{p-value} & \textbf{Adjusted Estimate} & \textbf{SE} & 
\textbf{p-value}\\
\midrule 
Semi-parametric Cox & -0.173 & 0.084 & 0.035 & -0.216 & 0.084 &  0.010 \\
Weibull PHM & -0.178 & 0.084 & 0.033 & -0.220 & 0.084 & 0.009\\
Semi-parametric frailty &  -0.235 & 0.098 & 0.017 & -0.216 & 0.084 & 0.010\\
Weibull frailty & -0.233 & 0.113 & 0.039 & -0.224 & 0.092 & 0.015\\
AFT EV & 0.110 & 0.052 & 0.033 &  0.133 & 0.051 & 0.009 \\
AFT Lognormal & 0.167 & 0.061 & 0.006 & 0.163 & 0.060 & 0.006 \\
AFT Loglogistic & 0.158 & 0.059 & 0.008 & 0.166 & 0.058 & 0.004 \\
AFT Splines & 0.124 & 0.059 & 0.034 & 0.109  & 0.057 & 0.055 \\
\bottomrule
\end{tabular}}
\endgroup{}
\end{table*}

\vspace{4em}

\begin{table*}[h] 
\centering
\caption{Models' estimates of the marginal survival differences between treatment groups on the RTOG9202 dataset, estimated using either the Kaplan-Meier estimator or a Cox model with a time-dependent effect of the treatment, without adjustment on baseline covariates, for the three time points $t_{25} = 2.8, t_{50} = 5.3$ and $t_{75} = 7.8$ corresponding respectively to the 25th, 50th, and 75th percentiles of the observed event times. For the Kaplan–Meier-based estimator, asymptotic standard deviations were computed using the Greenwood formula. For the estimator based on the Cox model with a time-dependent effect for the exposure, asymptotic standard deviations were obtained via non-parametric bootstrap using B $=500$ resampled datasets.} 
\label{tab:appli-surv} 
\vspace{0.3cm}
\begingroup
\fontsize{8}{10}\selectfont

\resizebox{\ifdim\width>\linewidth\linewidth\else\width\fi}{!}{
\begin{tabular}{l r r r r}
\toprule
\textbf{Model} & \textbf{Estimand} &\textbf{Estimation} & \textbf{SE} & \textbf{p-value}\\
\midrule
  &  $S_{M}(t_{25}|X=1)-S_{M}(t_{25}|X=0)$ & 0.067 & 0.020 & 0.001 \\
\cmidrule{2-5}
   & $S_{M}(t_{50}|X=1)-S_{M}(t_{50}|X=0)$ & 0.063 & 0.027 & 0.020 \\
\cmidrule{2-5}
 \multirow{-3}{*}{\raggedright\arraybackslash Kaplan-Meier} & $S_{M}(t_{75}|X=1)-S_{M}(t_{75}|X=0)$ & 0.077 & 0.031 & 0.013 \\
\cmidrule{1-5}
  & $S_{M}(t_{25}|X=1)-S_{M}(t_{25}|X=0)$ & 0.060 & 0.020 & 0.002 \\
\cmidrule{2-5}
   & $S_{M}(t_{50}|X=1)-S_{M}(t_{50}|X=0)$ & 0.074 & 0.026 & 0.004 \\
\cmidrule{2-5}
\multirow{-3}{*}{\raggedright\arraybackslash Time dependent Cox} & $S_{M}(t_{75}|X=1)-S_{M}(t_{75}|X=0)$ & 0.072 & 0.030 & 0.016 \\
\bottomrule
\end{tabular}}
\endgroup{}
\end{table*}

\clearpage

\bibliographystyle{unsrtnat}

\clearpage
\onecolumn
\setcounter{secnumdepth}{1}
\appendix
\renewcommand{\thesection}{Appendix~\Alph{section}}

\section{Proof of \eqref{eq:expecineq}}
\label{sec:expectations-inequality}

We first state the following preliminary result

\begin{align*}
R(u)&=\frac{f_{U\mid T>t, X=1}(u)}{f_{U\mid T>t, X=0}(u)} \text{ is increasing (resp.\ decreasing)} \\
&\implies
\mathbb{P}(U>u\mid T>t, X=1) \;\ge\; \mathbb{P}(U>u\mid T>t, X=0) \quad (\text{resp.\ }\le)\;  \quad \forall u \\
&\iff\ \ \text{for every increasing } g:\enspace
\mathbb{E}_{U\mid T>t, X=1}[g(U)] \;\ge\; \mathbb{E}_{U\mid T>t, X=0}[g(U)] \quad (\text{resp.\ }\le)\\
&\iff\ \ \text{for every decreasing } g:\enspace
\mathbb{E}_{U\mid T>t, X=1}[g(U)] \;\le\; \mathbb{E}_{U\mid T>t, X=0}[g(U)] \quad (\text{resp.\ }\ge)
\end{align*}

where $f_{U\mid T>t, X=x}(.)$ denotes the density of $U$ given $T>t$ and $X=x$, and $\mathbb{E}_{U\mid T>t, X=x}[g(U)] $ denotes the expectation of $g(U)$ with respect to the distribution of $U$ conditionally on $T>t$ and $X=x$. A proof of the first implication is provided in Theorem 1.C.1 of Shaked and Shanthikumar (2007)\cite{shaked2007stochastic}. The first equivalence follows from Equation 1.A.7 in the same reference.  The second equivalence is immediate from the first by replacing $g$ with $-g$, noting that if $g$ is strictly decreasing, then $-g$ is strictly increasing. Theorem 1.A.8 in Shaked and Shanthikumar (2007)\cite{shaked2007stochastic} further shows that, in order to have strict inequalities for the expectations, $g$ must be strictly monotone and the distributions of $U \mid T>t,X=1$ and $U \mid T>t,X=0$ must differ, or equivalently $R(u)$ must be strictly monotone.\\

We have 

\begin{align*}
R(u) &= \frac{f_{U\mid T>t, X=1}(u)}{f_{U\mid T>t, X=0}(u)}\\
&=\left.
\dfrac{S_{C}(t \mid X=1,U=u)f_{U\mid X=1}(u)}{S_{M}(t \mid X=1)}
\ \middle/\ 
\dfrac{S_{C}(t \mid X=0,U=u)f_{U\mid X=0}(u)}{S_{M}(t \mid X=0)}
\right.\\
\end{align*}

This expression may be simplified since $f_{U\mid X=1}(u)=f_{U\mid X=0}(u)$ because $X \perp U$, and rewritten as:

\begin{align*}
R(u) &=\dfrac{S_{C}(t \mid X=1,U=u)}{S_{C}(t \mid X=0,U=u)}\dfrac{\int_{-\infty}^{+\infty}S_{C}(t \mid X=0,U=u)f_U(u)du}{\int_{-\infty}^{+\infty}S_{C}(t \mid X=1,U=u)f_U(u)du}
\end{align*}

The monotonicity of $R(u)$ depends only on the term $Q(u)=\dfrac{S_{C}(t \mid X=1,U=u)}{S_{C}(t \mid X=0,U=u)}$

Assuming that model \eqref{eq:true model} holds, we have

\begin{align*}
Q(u)&= \dfrac{\exp\left[-A_{0,C}(t)\exp(\beta_{C}+\beta_U u)\right]}{\exp\left[-A_{0,C}(t)\exp(\beta_U u)\right]}\\
&=\exp\left[-A_{0,C}(t)\exp\left(\beta_U u\right)\left(\exp(\beta_{C})-1\right)\right]
\end{align*}

Taking the derivative we have

\begin{align*}
\dfrac{\partial Q(u)}{\partial u}&=-A_{0,C}(t)\left(\exp(\beta_{C})-1\right)\beta_U\exp(\beta_U u)\exp\left[-A_{0,C}(t)\exp\left(\beta_U u\right)\left(\exp(\beta_{C})-1\right)\right]
\end{align*}

If $\beta_U>0$ (resp. $\beta_U<0$):\\

The function $g(u)=\exp(\beta_U u)$ is strictly increasing (resp. strictly decreasing).\\

If $\beta_{C}>0$ then $\dfrac{\partial Q(u)}{\partial u}<0$ (resp. $\dfrac{\partial Q(u)}{\partial u}>0$),  thus $R(u)$ is strictly decreasing (resp. strictly increasing).

If $\beta_{C}<0$ then $\dfrac{\partial Q(u)}{\partial u}>0$ (resp. $\dfrac{\partial Q(u)}{\partial u}<0$),  thus $R(u)$ is strictly increasing (resp. strictly decreasing). \\

As a consequence, if $\beta_{C}\neq 0$ and $\beta_U\neq 0$, the preliminary result can be applied showing that:  $$\mathbb{E}_{U \mid T>t,\, X=1}\big[\exp(\beta_U U)\big]
    \neq
    \mathbb{E}_{U \mid T>t,\, X=0}\big[\exp(\beta_U U)\big]
    $$

\qed

\setcounter{secnumdepth}{2}

\section{Proof of formula \eqref{eq:analytic hazard function}}
\label{sec:proof-analytic-hazard}
From \eqref{eq:marginalhazard}, $\forall t>0$:

\begin{equation*}
\lambda_M(t\mid X=x)=\lambda_{0,C}(t)\exp(\beta_{C}x)\mathbb{E}_{U\mid T>t,X=x}\left(\exp(\beta_U U)\right)
\end{equation*}

In the special case of the gamma frailty model, $\beta_U=1$ and $U \sim$ log-gamma($1/\theta$ , $1/\theta$) with a density function $f_U(u)=\dfrac{1}{\theta^{1/\theta}\Gamma(1/\theta)}\exp(u/\theta)\exp(-\exp(u)/\theta)$, where $\Gamma(1/\theta)=\int_0^{+\infty}x^{\frac{1}{\theta}-1}\exp(-x)dx$, thus:

\begin{align*}
\lambda_M(t\mid X=x)
&=\lambda_{0,C}(t)\exp(\beta_{C}x)\int_{-\infty}^{+\infty}\exp(u)f_{U \mid T>t,X=x}(u)du\\
&=\lambda_{0,C}(t)\exp(\beta_{C}x)\int_{-\infty}^{+\infty}\exp(u)\dfrac{S_{C}(t \mid X=x,U=u)f_U(u)}{\int_{-\infty}^{+\infty}S_{C}(t \mid X=x,U=u)f_U(u)du}du\\
\end{align*}

We have

\begin{align*}
&\int_{-\infty}^{+\infty}S_{C}(t \mid X=x,U=u)f_U(u)du \\
&=\int_{-\infty}^{+\infty}\exp\left[-A_{0,C}(t)\exp(\beta_{C}x)\exp(u)\right]\dfrac{1}{\theta^{1/\theta}\Gamma(1/\theta)}\exp(u/\theta)\exp(-\exp(u)/\theta)du
\end{align*}

Let $u=\log(v)$, then

\begin{align*}
&\int_{-\infty}^{+\infty}S_{C}(t \mid X=x,U=u)f_U(u)du =\int_{0}^{+\infty}\exp\left[-A_{0,C}(t)\exp(\beta_{C}x)v\right]\dfrac{1}{\theta^{1/\theta}\Gamma(1/\theta)}v^{1/\theta}\exp(-v/\theta)\dfrac{dv}{v}\\
&=\dfrac{1}{\theta^{1/\theta}\Gamma(1/\theta)}\int_{0}^{+\infty}v^{\frac{1}{\theta}-1}\exp\left[-v\left(\dfrac{1}{\theta}+A_{0,C}(t)\exp(\beta_{C}x)\right)\right]dv\\
&=\frac{\displaystyle \int_{0}^{\infty}
  \dfrac{\left(\tfrac{1}{\theta}+A_{0,C}(t)\exp(\beta_{C}x)\right)^{1/\theta}}{\Gamma(1/\theta)}
  v^{\frac{1}{\theta}-1}
  \exp\!\left[-v\!\left(\tfrac{1}{\theta}+A_{0,C}(t)\exp(\beta_{C}x)\right)\right] \, dv}
{\theta^{1/\theta}\left(\tfrac{1}{\theta}+A_{0,C}(t)\exp(\beta_{C}x)\right)^{1/\theta}}\\
\end{align*}

The term in the numerator is the integral of the density of a 
Gamma $\!\left(\tfrac{1}{\theta},\, \tfrac{1}{\theta}+A_{0,C}(t)\exp(\beta_{C}x)\right)$ 
distribution, hence it equals $1$. This gives us

\begin{align*}
\int_{-\infty}^{+\infty}S_{C}(t \mid X=x,U=u)f_U(u)du
&=\dfrac{1}{\theta^{1/\theta}\left(\dfrac{1}{\theta}+A_{0,C}(t)\exp(\beta_{C}x)\right)^{1/\theta}}\\
&=\dfrac{1}{\left(1+\theta A_{0,C}(t)\exp(\beta_{C}x)\right)^{1/\theta}}
\end{align*}

Thus

\resizebox{\textwidth}{!}{%
\begin{minipage}{\textwidth}
\begin{align*}
\lambda_M(t\mid X=x)
&= \lambda_{0,C}(t)\exp(\beta_{C}x)\int_{-\infty}^{+\infty} 
\exp(u)\left(1+\theta A_{0,C}(t)\exp(\beta_{C}x)\right)^{1/\theta}\exp\left[-A_{0,C}(t)\exp(\beta_{C}x)\exp(u)\right] \\
&\quad \times \frac{1}{\theta^{1/\theta}\Gamma(1/\theta)}
\exp(u/\theta)\exp\!\big(-\exp(u)/\theta\big)\, du\\
\end{align*}
\end{minipage}%
}

Let $u=\log(v)$, then

\resizebox{\textwidth}{!}{%
\begin{minipage}{\textwidth}
\begin{align*}
\lambda_M(t\mid X=x)
&= \frac{\lambda_{0,C}(t)\exp(\beta_{C}x)\left(1+\theta A_{0,C}(t)\exp(\beta_{C}x)\right)^{1/\theta}}{\theta^{1/\theta}\Gamma(1/\theta)}\int_{0}^{+\infty} 
v^{1/\theta}\exp\left[-v\left(\dfrac{1}{\theta}+A_{0,C}(t)\exp(\beta_{C}x)\right)\right]dv\\
&= \frac{\lambda_{0,C}(t)\exp(\beta_{C}x)\left(1+\theta A_{0,C}(t)\exp(\beta_{C}x)\right)^{1/\theta}\Gamma\left(\frac{1}{\theta}+1\right)}{\left(\frac{1}{\theta}+A_{0,C}(t)\exp(\beta_{C}x)\right)^{\frac{1}{\theta}+1}\theta^{1/\theta}\Gamma(1/\theta)}\\
&\quad \times \int_{0}^{+\infty} \frac{\left(\frac{1}{\theta}+A_{0,C}(t)\exp(\beta_{C}x)\right)^{\frac{1}{\theta}+1}}{\Gamma\left(\frac{1}{\theta}+1\right)}v^{1/\theta}\exp\left[-v\left(\dfrac{1}{\theta}+A_{0,C}(t)\exp(\beta_{C}x)\right)\right]dv\\
\end{align*}
\end{minipage}%
}

The term inside the integral is the density of a Gamma $\!\left(\tfrac{1}{\theta}+1,\, \tfrac{1}{\theta}+A_{0,C}(t)\exp(\beta_{C}x)\right)$ 
distribution, hence the integral is equal to $1$. 

Thus 

\resizebox{\textwidth}{!}{%
\begin{minipage}{\textwidth}
\begin{align*}
\lambda_M(t\mid X=x)
&= \frac{\lambda_{0,C}(t)\exp(\beta_{C}x)\left(1+\theta A_{0,C}(t)\exp(\beta_{C}x)\right)^{1/\theta}\Gamma\left(\frac{1}{\theta}+1\right)}{\left(\frac{1}{\theta}+A_{0,C}(t)\exp(\beta_{C}x)\right)^{\frac{1}{\theta}+1}\theta^{1/\theta}\Gamma(1/\theta)}\\
&=\frac{\lambda_{0,C}(t)\exp(\beta_{C}x)\left(1+\theta A_{0,C}(t)\exp(\beta_{C}x)\right)^{1/\theta}}{\left(1+\theta A_{0,C}(t)\exp(\beta_{C}x)\right)^{\frac{1}{\theta}+1}}
\end{align*}
\end{minipage}%
}

Which gives

\begin{equation*}
\lambda_M(t|X = x) = \frac{\lambda_{0,C}(t) \exp(\beta_{C}x)}{1 + \theta A_{0,C}(t) \exp(\beta_{C}x)}
\end{equation*}

\qed

\clearpage

\makeatletter
\newcommand{\distas}[1]{\mathbin{\overset{#1}{\kern\z@\sim}}}%
\newsavebox{\mybox}\newsavebox{\mysim}
\newcommand{\distras}[1]{%
  \savebox{\mybox}{\hbox{\kern3pt$\scriptstyle#1$\kern3pt}}%
  \savebox{\mysim}{\hbox{$\sim$}}%
  \mathbin{\overset{#1}{\kern\z@\resizebox{\wd\mybox}{\ht\mysim}{$\sim$}}}%
}
\makeatother

\newcommand{\mytexttilde}{\raisebox{0.5ex}{\texttildelow}}
\graphicspath{{images/}}

\begin{center}
\Huge{Supplementary material for \\ 
"Built-in Selection Bias in Proportional Hazards Models with Omitted Covariates: Simulation Evidence and Alternative Approaches"}
\end{center}

\maketitle

\clearpage

\begin{table*}[h!]
\centering
\renewcommand{\thetable}{S1}
\caption{Results of the simulation study with $U \sim \mathcal{N}(0,1)$ and $\beta_U \in \{0.2, 0.4, 0.8, 1\}$. Are presented the true value of the regression parameter for the exposure, its estimate, bias, standardized bias, asymptotic and empirical standard deviations and the coverage rate of the $95\%$ confidence interval (and estimated variance of the frailty for the frailty model). All results are based on $1,000$ simulated replicates.\\}
\label{tab:results_U_gaussian}

\begingroup
\fontsize{8}{10}\selectfont
\resizebox{\textwidth}{!}{
\begin{tabular}{rllllllllll}
\toprule
\textbf{{$\beta_U$}} & \textbf{Model} & \textbf{Estimand} & \textbf{True} & \textbf{Estimation} & \textbf{Bias} & \textbf{${\hat{\sigma}_{as}}$} & \textbf{${\hat{\sigma}_{emp}}$} & \textbf{Standardized bias} & \textbf{$\hat{\theta}$} & \textbf{Coverage Rate}\\
\midrule
 & Semi-parametric Cox & ${\beta}_{adj}$ & -0.6 & -0.587 & 0.013 & 0.092 & 0.094 & 13.59\% &  & 94.1\%\\
\cmidrule{2-11}
 & Weibull PHM & ${\beta}_{adj}$ & -0.6 & -0.590 & 0.010 & 0.092 & 0.093 & 11.28\% &  & 94.2\%\\
\cmidrule{2-11}
 & Semi-parametric frailty & ${\beta}_{adj}$ & -0.6 & -0.601 & -0.001 & 0.094 & 0.104 & 1.02\% & 3.603 & 92.8\%\\
\cmidrule{2-11}
 & Weibull frailty & ${\beta}_{adj}$ & -0.6 & -0.606 & -0.006 & 0.098 & 0.097 & 5.94\% & 0.066 & 95.2\%\\
\cmidrule{2-11}
 & AFT EV & $\frac{{\beta}_{adj}}{b}$ & -0.066 & -0.066 & 0.000 & 0.010 & 0.010 & 3.80\% &  & 94.5\%\\
\cmidrule{2-11}
 & AFT Lognormal & $\frac{{\beta}_{adj}}{b}$ & -0.066 & -0.070 & -0.003 & 0.013 & 0.013 & 26.15\% &  & 94.8\%\\
\cmidrule{2-11}
 & AFT Loglogistic & $\frac{{\beta}_{adj}}{b}$ & -0.066 & -0.068 & -0.001 & 0.012 & 0.011 & 8.10\% &  & 95.4\%\\
\cmidrule{2-11}
\multirow{-8}{*}[3\dimexpr\aboverulesep+\belowrulesep+\cmidrulewidth]{\raggedleft\arraybackslash 0.2} & AFT Splines & $\frac{{\beta}_{adj}}{b}$ & -0.066 & -0.066 & 0.000 & 0.010 & 0.010 & 2.32\% &  & 94.6\%\\
\cmidrule{1-11}
\addlinespace[2em]
\hspace{1em} & Semi-parametric Cox & ${\beta}_{adj}$ & -0.6 & -0.556 & 0.044 & 0.092 & 0.092 & 47.44\% &  & 92.8\%\\
\cmidrule{2-11}
\hspace{1em} & Weibull PHM & ${\beta}_{adj}$ & -0.6 & -0.564 & 0.036 & 0.092 & 0.093 & 39.28\% &  & 93.0\%\\
\cmidrule{2-11}
 & Semi-parametric frailty & ${\beta}_{adj}$ & -0.6 & -0.579 & 0.021 & 0.095 & 0.109 & 19.32\% & 3.157 & 92.4\%\\
\cmidrule{2-11}
 & Weibull frailty & ${\beta}_{adj}$ & -0.6 & -0.599 & 0.001 & 0.103 & 0.100 & 1.17\% & 0.147 & 96.0\%\\
\cmidrule{2-11}
 & AFT EV & $\frac{{\beta}_{adj}}{b}$ & -0.066 & -0.066 & 0.001 & 0.011 & 0.011 & 8.72\% &  & 94.2\%\\
\cmidrule{2-11}
 & AFT Lognormal & $\frac{{\beta}_{adj}}{b}$ & -0.066 & -0.070 & -0.003 & 0.013 & 0.013 & 24.77\% &  & 94.6\%\\
\cmidrule{2-11}
 & AFT Loglogistic & $\frac{{\beta}_{adj}}{b}$ & -0.066 & -0.068 & -0.001 & 0.012 & 0.012 & 7.83\% &  & 95.6\%\\
\cmidrule{2-11}
\multirow{-8}{*}[3\dimexpr\aboverulesep+\belowrulesep+\cmidrulewidth]{\raggedleft\arraybackslash 0.4} & AFT Splines & $\frac{{\beta}_{adj}}{b}$ & -0.066 & -0.066 & 0.000 & 0.011 & 0.011 & 2.88\% &  & 95.0\%\\
\cmidrule{1-11}
\addlinespace[2em]
\hspace{1em} & Semi-parametric Cox & ${\beta}_{adj}$ & -0.6 & -0.477 & 0.123 & 0.092 & 0.091 & 134.74\% &  & 73.3\%\\
\cmidrule{2-11}
\hspace{1em} & Weibull PHM & ${\beta}_{adj}$ & -0.6 & -0.490 & 0.110 & 0.091 & 0.094 & 117.75\% &  & 76.5\%\\
\cmidrule{2-11}
 & Semi-parametric frailty & ${\beta}_{adj}$ & -0.6 & -0.523 & 0.077 & 0.098 & 0.129 & 59.53\% & 2.537 & 78.9\%\\
\cmidrule{2-11}
 & Weibull frailty & ${\beta}_{adj}$ & -0.6 & -0.582 & 0.018 & 0.116 & 0.114 & 15.54\% & 0.453 & 94.9\%\\
\cmidrule{2-11}
 & AFT EV & $\frac{{\beta}_{adj}}{b}$ & -0.066 & -0.064 & 0.003 & 0.012 & 0.012 & 21.11\% &  & 93.9\%\\
\cmidrule{2-11}
 & AFT Lognormal & $\frac{{\beta}_{adj}}{b}$ & -0.066 & -0.069 & -0.003 & 0.014 & 0.014 & 19.53\% &  & 94.0\%\\
\cmidrule{2-11}
 & AFT Loglogistic & $\frac{{\beta}_{adj}}{b}$ & -0.066 & -0.067 & -0.001 & 0.013 & 0.013 & 5.60\% &  & 95.1\%\\
\cmidrule{2-11}
\multirow{-8}{*}[3\dimexpr\aboverulesep+\belowrulesep+\cmidrulewidth]{\raggedleft\arraybackslash 0.8} & AFT Splines & $\frac{{\beta}_{adj}}{b}$ & -0.066 & -0.067 & 0.000 & 0.013 & 0.012 & 1.04\% &  & 94.7\%\\
\cmidrule{1-11}
\addlinespace[2em]
\hspace{1em} & Semi-parametric Cox & ${\beta}_{adj}$ & -0.6 & -0.440 & 0.160 & 0.092 & 0.090 & 178.47\% &  & 59.3\%\\
\cmidrule{2-11}
\hspace{1em} & Weibull PHM & ${\beta}_{adj}$ & -0.6 & -0.453 & 0.147 & 0.091 & 0.093 & 158.58\% &  & 63.8\%\\
\cmidrule{2-11}
 & Semi-parametric frailty & ${\beta}_{adj}$ & -0.6 & -0.493 & 0.107 & 0.100 & 0.132 & 80.48\% & 2.404 & 69.5\%\\
\cmidrule{2-11}
 & Weibull frailty & ${\beta}_{adj}$ & -0.6 & -0.569 & 0.031 & 0.121 & 0.119 & 26.25\% & 0.634 & 95.0\%\\
\cmidrule{2-11}
 & AFT EV & $\frac{{\beta}_{adj}}{b}$ & -0.066 & -0.063 & 0.003 & 0.013 & 0.013 & 25.21\% &  & 94.3\%\\
\cmidrule{2-11}
 & AFT Lognormal & $\frac{{\beta}_{adj}}{b}$ & -0.066 & -0.069 & -0.003 & 0.015 & 0.015 & 17.41\% &  & 94.6\%\\
\cmidrule{2-11}
 & AFT Loglogistic & $\frac{{\beta}_{adj}}{b}$ & -0.066 & -0.067 & -0.001 & 0.014 & 0.014 & 4.71\% &  & 95.6\%\\
\cmidrule{2-11}
\multirow{-8}{*}[3\dimexpr\aboverulesep+\belowrulesep+\cmidrulewidth]{\raggedleft\arraybackslash 1.0} & AFT Splines & $\frac{{\beta}_{adj}}{b}$ & -0.066 & -0.067 & 0.000 & 0.014 & 0.013 & 1.29\% &  & 96.1\%\\
\bottomrule
\end{tabular}}
\endgroup
\end{table*}

\clearpage

\begin{table*}[h!]
\centering
\renewcommand{\thetable}{S2}
\caption{Results of the simulation study with $U \sim$ log-gamma(1,1) and $\beta_U \in \{0.2, 0.4, 0.8, 1\}$. Are presented the true value of the regression parameter for the exposure, its estimate, bias, standardized bias, asymptotic and empirical standard deviations and the coverage rate of the $95\%$ confidence interval (and estimated variance of the frailty for the frailty model). All results are based on $1,000$ simulated replicates.\\}
\label{tab:results_U_loggamma}

\begingroup
\fontsize{8}{10}\selectfont
\resizebox{\textwidth}{!}{
\begin{tabular}{rllllllllll}
\toprule
\textbf{{$\beta_U$}} & \textbf{Model} & \textbf{Estimand} & \textbf{True} & \textbf{Estimation} & \textbf{Bias} & \textbf{${\hat{\sigma}_{as}}$} & \textbf{${\hat{\sigma}_{emp}}$} & \textbf{Standardized bias} & \textbf{$\hat{\theta}$} & \textbf{Coverage Rate}\\
\midrule
 & Semi-parametric Cox & ${\beta}_{adj}$ & -0.6 & -0.584 & 0.016 & 0.092 & 0.094 & 16.49\% &  & 95.1\%\\
\cmidrule{2-11}
 & Weibull PHM & ${\beta}_{adj}$ & -0.6 & -0.588 & 0.012 & 0.091 & 0.094 & 12.88\% &  & 95.2\%\\
\cmidrule{2-11}
 & Semi-parametric frailty & ${\beta}_{adj}$ & -0.6 & -0.605 & -0.005 & 0.095 & 0.110 & 4.57\% & 3.177 & 92.9\%\\
\cmidrule{2-11}
 & Weibull frailty & ${\beta}_{adj}$ & -0.6 & -0.613 & -0.013 & 0.100 & 0.100 & 12.50\% & 0.090 & 95.3\%\\
\cmidrule{2-11}
 & AFT EV & $\frac{{\beta}_{adj}}{b}$ & -0.066 & -0.066 & 0.000 & 0.010 & 0.011 & 1.69\% &  & 94.8\%\\
\cmidrule{2-11}
 & AFT Lognormal & $\frac{{\beta}_{adj}}{b}$ & -0.066 & -0.071 & -0.004 & 0.013 & 0.013 & 33.90\% &  & 93.3\%\\
\cmidrule{2-11}
 & AFT Loglogistic & $\frac{{\beta}_{adj}}{b}$ & -0.066 & -0.068 & -0.002 & 0.012 & 0.011 & 14.89\% &  & 94.9\%\\
\cmidrule{2-11}
\multirow{-8}{*}[3\dimexpr\aboverulesep+\belowrulesep+\cmidrulewidth]{\raggedleft\arraybackslash 0.2} & AFT Splines & $\frac{{\beta}_{adj}}{b}$ & -0.066 & -0.067 & 0.000 & 0.010 & 0.011 & 0.85\% &  & 95.2\%\\
\cmidrule{1-11}
\addlinespace[2em]
\hspace{1em} & Semi-parametric Cox & ${\beta}_{adj}$ & -0.6 & -0.531 & 0.069 & 0.094 & 0.090 & 76.08\% &  & 89.5\%\\
\cmidrule{2-11}
\hspace{1em} & Weibull PHM & ${\beta}_{adj}$ & -0.6 & -0.541 & 0.059 & 0.093 & 0.092 & 64.55\% &  & 90.7\%\\
\cmidrule{2-11}
 & Semi-parametric frailty & ${\beta}_{adj}$ & -0.6 & -0.569 & 0.031 & 0.099 & 0.121 & 25.74\% & 2.618 & 89.9\%\\
\cmidrule{2-11}
 & Weibull frailty & ${\beta}_{adj}$ & -0.6 & -0.597 & 0.003 & 0.109 & 0.107 & 3.08\% & 0.241 & 95.4\%\\
\cmidrule{2-11}
 & AFT EV & $\frac{{\beta}_{adj}}{b}$ & -0.066 & -0.064 & 0.003 & 0.011 & 0.011 & 23.57\% &  & 94.6\%\\
\cmidrule{2-11}
 & AFT Lognormal & $\frac{{\beta}_{adj}}{b}$ & -0.066 & -0.069 & -0.002 & 0.013 & 0.013 & 17.96\% &  & 95.3\%\\
\cmidrule{2-11}
 & AFT Loglogistic & $\frac{{\beta}_{adj}}{b}$ & -0.066 & -0.067 & 0.000 & 0.012 & 0.012 & 0.54\% &  & 96.1\%\\
\cmidrule{2-11}
\multirow{-8}{*}[3\dimexpr\aboverulesep+\belowrulesep+\cmidrulewidth]{\raggedleft\arraybackslash 0.4} & AFT Splines & $\frac{{\beta}_{adj}}{b}$ & -0.066 & -0.065 & 0.001 & 0.011 & 0.011 & 13.39\% &  & 95.9\%\\
\cmidrule{1-11}
\addlinespace[2em]
\hspace{1em} & Semi-parametric Cox & ${\beta}_{adj}$ & -0.6 & -0.453 & 0.147 & 0.097 & 0.094 & 157.20\% &  & 67.8\%\\
\cmidrule{2-11}
\hspace{1em} & Weibull PHM & ${\beta}_{adj}$ & -0.6 & -0.463 & 0.137 & 0.097 & 0.098 & 139.88\% &  & 70.9\%\\
\cmidrule{2-11}
 & Semi-parametric frailty & ${\beta}_{adj}$ & -0.6 & -0.530 & 0.070 & 0.109 & 0.150 & 46.63\% & 2.056 & 77.8\%\\
\cmidrule{2-11}
 & Weibull frailty & ${\beta}_{adj}$ & -0.6 & -0.596 & 0.004 & 0.130 & 0.127 & 2.97\% & 0.730 & 95.3\%\\
\cmidrule{2-11}
 & AFT EV & $\frac{{\beta}_{adj}}{b}$ & -0.066 & -0.061 & 0.005 & 0.013 & 0.013 & 40.92\% &  & 92.7\%\\
\cmidrule{2-11}
 & AFT Lognormal & $\frac{{\beta}_{adj}}{b}$ & -0.066 & -0.068 & -0.001 & 0.015 & 0.014 & 9.23\% &  & 94.5\%\\
\cmidrule{2-11}
 & AFT Loglogistic & $\frac{{\beta}_{adj}}{b}$ & -0.066 & -0.066 & 0.001 & 0.014 & 0.013 & 5.12\% &  & 96.1\%\\
\cmidrule{2-11}
\multirow{-8}{*}[3\dimexpr\aboverulesep+\belowrulesep+\cmidrulewidth]{\raggedleft\arraybackslash 0.8} & AFT Splines & $\frac{{\beta}_{adj}}{b}$ & -0.066 & -0.065 & 0.001 & 0.014 & 0.013 & 10.36\% &  & 95.5\%\\
\cmidrule{1-11}
\addlinespace[2em]
\hspace{1em} & Semi-parametric Cox & ${\beta}_{adj}$ & -0.6 & -0.420 & 0.180 & 0.098 & 0.095 & 189.01\% &  & 55.9\%\\
\cmidrule{2-11}
\hspace{1em} & Weibull PHM & ${\beta}_{adj}$ & -0.6 & -0.429 & 0.171 & 0.098 & 0.100 & 171.06\% &  & 59.7\%\\
\cmidrule{2-11}
 & Semi-parametric frailty & ${\beta}_{adj}$ & -0.6 & -0.509 & 0.091 & 0.113 & 0.160 & 56.96\% & 1.983 & 71.8\%\\
\cmidrule{2-11}
 & Weibull frailty & ${\beta}_{adj}$ & -0.6 & -0.592 & 0.008 & 0.139 & 0.136 & 6.16\% & 1.014 & 95.0\%\\
\cmidrule{2-11}
 & AFT EV & $\frac{{\beta}_{adj}}{b}$ & -0.066 & -0.060 & 0.007 & 0.014 & 0.014 & 46.83\% &  & 91.8\%\\
\cmidrule{2-11}
 & AFT Lognormal & $\frac{{\beta}_{adj}}{b}$ & -0.066 & -0.067 & -0.001 & 0.016 & 0.015 & 4.61\% &  & 95.1\%\\
\cmidrule{2-11}
 & AFT Loglogistic & $\frac{{\beta}_{adj}}{b}$ & -0.066 & -0.066 & 0.001 & 0.015 & 0.014 & 8.09\% &  & 95.5\%\\
\cmidrule{2-11}
\multirow{-8}{*}[3\dimexpr\aboverulesep+\belowrulesep+\cmidrulewidth]{\raggedleft\arraybackslash 1.0} & AFT Splines & $\frac{{\beta}_{adj}}{b}$ & -0.066 & -0.066 & 0.001 & 0.015 & 0.014 & 7.87\% &  & 95.5\%\\\\
\bottomrule
\end{tabular}}
\endgroup
\end{table*}

\clearpage
\begin{table*}[h!]
\centering
\renewcommand{\thetable}{S3}
\caption{Results of the simulation study with $U \sim \mathcal{B}er(0.5)$ and $\beta_U \in \{0.2, 0.4, 0.8, 1\}$. Are presented the true value of the regression parameter for the exposure, its estimate, bias, standardized bias, asymptotic and empirical standard deviations and the coverage rate of the $95\%$ confidence interval (and estimated variance of the frailty for the frailty model). All results are based on $1,000$ simulated replicates.\\}
\label{tab:results_U_binaire}

\begingroup
\fontsize{8}{10}\selectfont
\resizebox{\textwidth}{!}{
\begin{tabular}{rllllllllll}
\toprule
\textbf{{$\beta_U$}} & \textbf{Model} & \textbf{Estimand} & \textbf{True} & \textbf{Estimation} & \textbf{Bias} & \textbf{${\hat{\sigma}_{as}}$} & \textbf{${\hat{\sigma}_{emp}}$} & \textbf{Standardized bias} & \textbf{$\hat{\theta}$} & \textbf{Coverage Rate}\\
\midrule
 & Semi-parametric Cox & ${\beta}_{adj}$ & -0.6 & -0.598 & 0.002 & 0.100 & 0.101 & 1.53\% &  & 95.2\%\\
\cmidrule{2-11}
 & Weibull PHM & ${\beta}_{adj}$ & -0.6 & -0.599 & 0.001 & 0.099 & 0.101 & 0.60\% &  & 95.4\%\\
\cmidrule{2-11}
 & Semi-parametric frailty & ${\beta}_{adj}$ & -0.6 & -0.614 & -0.014 & 0.102 & 0.113 & 12.09\% & 3.668 & 93.5\%\\
\cmidrule{2-11}
 & Weibull frailty & ${\beta}_{adj}$ & -0.6 & -0.616 & -0.016 & 0.106 & 0.105 & 14.83\% & 0.071 & 95.8\%\\
\cmidrule{2-11}
 & AFT EV & $\frac{{\beta}_{adj}}{b}$ & -0.066 & -0.067 & 0.000 & 0.011 & 0.011 & 0.46\% &  & 94.9\%\\
\cmidrule{2-11}
 & AFT Lognormal & $\frac{{\beta}_{adj}}{b}$ & -0.066 & -0.071 & -0.004 & 0.013 & 0.014 & 31.90\% &  & 93.7\%\\
\cmidrule{2-11}
 & AFT Loglogistic & $\frac{{\beta}_{adj}}{b}$ & -0.066 & -0.068 & -0.002 & 0.012 & 0.012 & 12.55\% &  & 94.8\%\\
\cmidrule{2-11}
\multirow{-8}{*}[3\dimexpr\aboverulesep+\belowrulesep+\cmidrulewidth]{\raggedleft\arraybackslash 0.2} & AFT Splines & $\frac{{\beta}_{adj}}{b}$ & -0.066 & -0.067 & 0.000 & 0.011 & 0.011 & 1.59\% &  & 95.2\%\\
\cmidrule{1-11}
\addlinespace[2em]
\hspace{1em} & Semi-parametric Cox & ${\beta}_{adj}$ & -0.6 & -0.590 & 0.010 & 0.098 & 0.098 & 10.38\% &  & 95.1\%\\
\cmidrule{2-11}
\hspace{1em} & Weibull PHM & ${\beta}_{adj}$ & -0.6 & -0.592 & 0.008 & 0.097 & 0.098 & 8.16\% &  & 94.8\%\\
\cmidrule{2-11}
 & Semi-parametric frailty & ${\beta}_{adj}$ & -0.6 & -0.606 & -0.006 & 0.100 & 0.111 & 5.31\% & 3.580 & 93.2\%\\
\cmidrule{2-11}
 & Weibull frailty & ${\beta}_{adj}$ & -0.6 & -0.612 & -0.012 & 0.105 & 0.104 & 11.54\% & 0.086 & 95.2\%\\
\cmidrule{2-11}
 & AFT EV & $\frac{{\beta}_{adj}}{b}$ & -0.066 & -0.067 & 0.000 & 0.011 & 0.011 & 1.33\% &  & 95.0\%\\
\cmidrule{2-11}
 & AFT Lognormal & $\frac{{\beta}_{adj}}{b}$ & -0.066 & -0.071 & -0.004 & 0.013 & 0.013 & 30.25\% &  & 93.9\%\\
\cmidrule{2-11}
 & AFT Loglogistic & $\frac{{\beta}_{adj}}{b}$ & -0.066 & -0.068 & -0.001 & 0.012 & 0.012 & 11.30\% &  & 94.7\%\\
\cmidrule{2-11}
\multirow{-8}{*}[3\dimexpr\aboverulesep+\belowrulesep+\cmidrulewidth]{\raggedleft\arraybackslash 0.4} & AFT Splines & $\frac{{\beta}_{adj}}{b}$ & -0.066 & -0.067 & 0.000 & 0.011 & 0.011 & 0.91\% &  & 95.5\%\\
\cmidrule{1-11}
\addlinespace[2em]
\hspace{1em} & Semi-parametric Cox & ${\beta}_{adj}$ & -0.6 & -0.556 & 0.044 & 0.093 & 0.095 & 46.81\% &  & 91.6\%\\
\cmidrule{2-11}
\hspace{1em} & Weibull PHM & ${\beta}_{adj}$ & -0.6 & -0.563 & 0.037 & 0.092 & 0.095 & 38.77\% &  & 92.5\%\\
\cmidrule{2-11}
 & Semi-parametric frailty & ${\beta}_{adj}$ & -0.6 & -0.581 & 0.019 & 0.096 & 0.114 & 16.38\% & 3.043 & 90.6\%\\
\cmidrule{2-11}
 & Weibull frailty & ${\beta}_{adj}$ & -0.6 & -0.603 & -0.003 & 0.105 & 0.107 & 2.73\% & 0.167 & 95.1\%\\
\cmidrule{2-11}
 & AFT EV & $\frac{{\beta}_{adj}}{b}$ & -0.066 & -0.066 & 0.001 & 0.011 & 0.011 & 8.51\% &  & 94.7\%\\
\cmidrule{2-11}
 & AFT Lognormal & $\frac{{\beta}_{adj}}{b}$ & -0.066 & -0.070 & -0.003 & 0.013 & 0.013 & 24.99\% &  & 94.2\%\\
\cmidrule{2-11}
 & AFT Loglogistic & $\frac{{\beta}_{adj}}{b}$ & -0.066 & -0.068 & -0.001 & 0.012 & 0.012 & 8.10\% &  & 94.8\%\\
\cmidrule{2-11}
\multirow{-8}{*}[3\dimexpr\aboverulesep+\belowrulesep+\cmidrulewidth]{\raggedleft\arraybackslash 0.8} & AFT Splines & $\frac{{\beta}_{adj}}{b}$ & -0.066 & -0.066 & 0.000 & 0.011 & 0.011 & 1.61\% &  & 95.4\%\\
\cmidrule{1-11}
\addlinespace[2em]
\hspace{1em} & Semi-parametric Cox & ${\beta}_{adj}$ & -0.6 & -0.533 & 0.067 & 0.091 & 0.091 & 73.80\% &  & 86.9\%\\
\cmidrule{2-11}
\hspace{1em} & Weibull PHM & ${\beta}_{adj}$ & -0.6 & -0.543 & 0.057 & 0.090 & 0.093 & 61.57\% &  & 88.8\%\\
\cmidrule{2-11}
 & Semi-parametric frailty & ${\beta}_{adj}$ & -0.6 & -0.565 & 0.035 & 0.095 & 0.117 & 29.93\% & 2.736 & 87.0\%\\
\cmidrule{2-11}
 & Weibull frailty & ${\beta}_{adj}$ & -0.6 & -0.601 & -0.001 & 0.106 & 0.107 & 0.99\% & 0.238 & 94.8\%\\
\cmidrule{2-11}
 & AFT EV & $\frac{{\beta}_{adj}}{b}$ & -0.066 & -0.065 & 0.001 & 0.011 & 0.011 & 13.23\% &  & 95.2\%\\
\cmidrule{2-11}
 & AFT Lognormal & $\frac{{\beta}_{adj}}{b}$ & -0.066 & -0.070 & -0.003 & 0.013 & 0.013 & 22.11\% &  & 94.3\%\\
\cmidrule{2-11}
 & AFT Loglogistic & $\frac{{\beta}_{adj}}{b}$ & -0.066 & -0.067 & -0.001 & 0.012 & 0.012 & 6.54\% &  & 94.9\%\\
\cmidrule{2-11}
\multirow{-8}{*}[3\dimexpr\aboverulesep+\belowrulesep+\cmidrulewidth]{\raggedleft\arraybackslash 1.0} & AFT Splines & $\frac{{\beta}_{adj}}{b}$ & -0.066 & -0.066 & 0.000 & 0.011 & 0.011 & 2.46\% &  & 95.8\%\\
\bottomrule\\
\bottomrule
\end{tabular}}
\endgroup
\end{table*}

\clearpage

\begin{table*}[h!]
\centering
\renewcommand{\thetable}{S4}
\caption{Results of the simulation study with $U \sim \mathcal{N}(0,1)$ and $\beta_U \in \{0.2, 0.4, 0.8, 1\}$. Are presented the true value of the survival differences at each time, its estimate, bias, standardized bias, asymptotic and empirical standard deviations and the coverage rate of the $95\%$ confidence interval. All results are based on 200 simulated replicates.\\}
\label{tab:results_U_SURV}

\begingroup
\fontsize{8}{10}\selectfont
\resizebox{\textwidth}{!}{
\begin{tabular}{rlllllllll}
\toprule
\textbf{{$\beta_U$}} & \textbf{Model} & \textbf{Estimand} & \textbf{True} & \textbf{Estimation} & \textbf{Bias} & \textbf{${\hat{\sigma}_{as}}$} & \textbf{${\hat{\sigma}_{emp}}$} & \textbf{Standardized bias} & \textbf{Coverage Rate}\\
\midrule
 &  & $S_{M}(8|X=1)-S_{M}(8|X=0)$ & 0.055 & 0.053 & -0.002 & 0.036 & 0.036 & 6.31\% & 91.5\%\\
\cmidrule{3-10}
 &  & $S_{M}(9|X=1)-S_{M}(9|X=0)$ & 0.130 & 0.124 & -0.006 & 0.056 & 0.053 & 10.89\% & 96.5\%\\
\cmidrule{3-10}
 & \multirow{-3}{*}{\raggedright\arraybackslash Kaplan-Meier} & $S_{M}(10|X=1)-S_{M}(10|X=0)$ & 0.207 & 0.199 & -0.008 & 0.072 & 0.074 & 10.71\% & 93.5\%\\
\cmidrule{2-10}
 &  & $S_{M}(8|X=1)-S_{M}(8|X=0)$ & 0.055 & 0.053 & -0.002 & 0.033 & 0.033 & 6.89\% & 94.5\%\\
\cmidrule{3-10}
 &  & $S_{M}(9|X=1)-S_{M}(9|X=0)$ & 0.130 & 0.125 & -0.005 & 0.051 & 0.051 & 10.33\% & 96.0\%\\
\cmidrule{3-10}
\multirow{-6}{*}[0.5\dimexpr\aboverulesep+\belowrulesep+\cmidrulewidth]{\raggedleft\arraybackslash 0.2} & \multirow{-3}{*}{\raggedright\arraybackslash Time dependent Cox} & $S_{M}(10|X=1)-S_{M}(10|X=0)$ & 0.207 & 0.198 & -0.009 & 0.065 & 0.065 & 14.09\% & 94.0\%\\
\cmidrule{1-10}
\addlinespace[2em]
\multicolumn{10}{l}{\textbf{}}\\
\hspace{1em} &  & $S_{M}(8|X=1)-S_{M}(8|X=0)$ & 0.058 & 0.058 & 0.000 & 0.037 & 0.038 & 0.62\% & 94.0\%\\
\cmidrule{3-10}
\hspace{1em} &  & $S_{M}(9|X=1)-S_{M}(9|X=0)$ & 0.131 & 0.126 & -0.005 & 0.057 & 0.055 & 8.76\% & 94.0\%\\
\cmidrule{3-10}
 & \multirow{-3}{*}{\raggedright\arraybackslash Kaplan-Meier} & $S_{M}(10|X=1)-S_{M}(10|X=0)$ & 0.199 & 0.192 & -0.006 & 0.072 & 0.070 & 8.87\% & 96.0\%\\
\cmidrule{2-10}
 &  & $S_{M}(8|X=1)-S_{M}(8|X=0)$ & 0.058 & 0.057 & 0.000 & 0.034 & 0.034 & 1.09\% & 94.0\%\\
\cmidrule{3-10}
 &  & $S_{M}(9|X=1)-S_{M}(9|X=0)$ & 0.131 & 0.126 & -0.005 & 0.052 & 0.051 & 9.32\% & 94.0\%\\
\cmidrule{3-10}
\multirow{-6}{*}[0.5\dimexpr\aboverulesep+\belowrulesep+\cmidrulewidth]{\raggedleft\arraybackslash 0.4} & \multirow{-3}{*}{\raggedright\arraybackslash Time dependent Cox} & $S_{M}(10|X=1)-S_{M}(10|X=0)$ & 0.199 & 0.191 & -0.008 & 0.065 & 0.064 & 12.39\% & 96.0\%\\
\cmidrule{1-10}
\addlinespace[2em]
\multicolumn{10}{l}{\textbf{}}\\
\hspace{1em} &  & $S_{M}(8|X=1)-S_{M}(8|X=0)$ & 0.065 & 0.062 & -0.003 & 0.039 & 0.039 & 7.99\% & 94.0\%\\
\cmidrule{3-10}
\hspace{1em} &  & $S_{M}(9|X=1)-S_{M}(9|X=0)$ & 0.130 & 0.133 & 0.003 & 0.058 & 0.060 & 5.31\% & 93.5\%\\
\cmidrule{3-10}
 & \multirow{-3}{*}{\raggedright\arraybackslash Kaplan-Meier} & $S_{M}(10|X=1)-S_{M}(10|X=0)$ & 0.171 & 0.170 & -0.001 & 0.071 & 0.078 & 0.90\% & 92.0\%\\
\cmidrule{2-10}
 &  & $S_{M}(8|X=1)-S_{M}(8|X=0)$ & 0.065 & 0.063 & -0.002 & 0.036 & 0.037 & 5.37\% & 93.0\%\\
\cmidrule{3-10}
 &  & $S_{M}(9|X=1)-S_{M}(9|X=0)$ & 0.130 & 0.131 & 0.001 & 0.054 & 0.055 & 2.13\% & 95.0\%\\
\cmidrule{3-10}
\multirow{-6}{*}[0.5\dimexpr\aboverulesep+\belowrulesep+\cmidrulewidth]{\raggedleft\arraybackslash 0.8} & \multirow{-3}{*}{\raggedright\arraybackslash Time dependent Cox} & $S_{M}(10|X=1)-S_{M}(10|X=0)$ & 0.171 & 0.168 & -0.003 & 0.065 & 0.071 & 4.21\% & 92.0\%\\
\cmidrule{1-10}
\addlinespace[2em]
\multicolumn{10}{l}{\textbf{}}\\
\hspace{1em} &  & $S_{M}(8|X=1)-S_{M}(8|X=0)$ & 0.070 & 0.071 & 0.001 & 0.042 & 0.042 & 2.50\% & 94.5\%\\
\cmidrule{3-10}
\hspace{1em} &  & $S_{M}(9|X=1)-S_{M}(9|X=0)$ & 0.126 & 0.122 & -0.004 & 0.059 & 0.058 & 7.26\% & 94.5\%\\
\cmidrule{3-10}
 & \multirow{-3}{*}{\raggedright\arraybackslash Kaplan-Meier} & $S_{M}(10|X=1)-S_{M}(10|X=0)$ & 0.156 & 0.144 & -0.013 & 0.071 & 0.072 & 17.36\% & 94.0\%\\
\cmidrule{2-10}
 &  & $S_{M}(8|X=1)-S_{M}(8|X=0)$ & 0.070 & 0.070 & 0.000 & 0.039 & 0.040 & 0.96\% & 93.5\%\\
\cmidrule{3-10}
 &  & $S_{M}(9|X=1)-S_{M}(9|X=0)$ & 0.126 & 0.122 & -0.005 & 0.054 & 0.056 & 8.24\% & 92.5\%\\
\cmidrule{3-10}
\multirow{-6}{*}[0.5\dimexpr\aboverulesep+\belowrulesep+\cmidrulewidth]{\raggedleft\arraybackslash 1.0} & \multirow{-3}{*}{\raggedright\arraybackslash Time dependent Cox} & $S_{M}(10|X=1)-S_{M}(10|X=0)$ & 0.156 & 0.144 & -0.012 & 0.065 & 0.068 & 18.35\% & 92.0\%\\
\bottomrule
\end{tabular}}
\endgroup
\end{table*}

\clearpage

\begin{figure*}
 \centering
  \includegraphics[width=16.8cm]{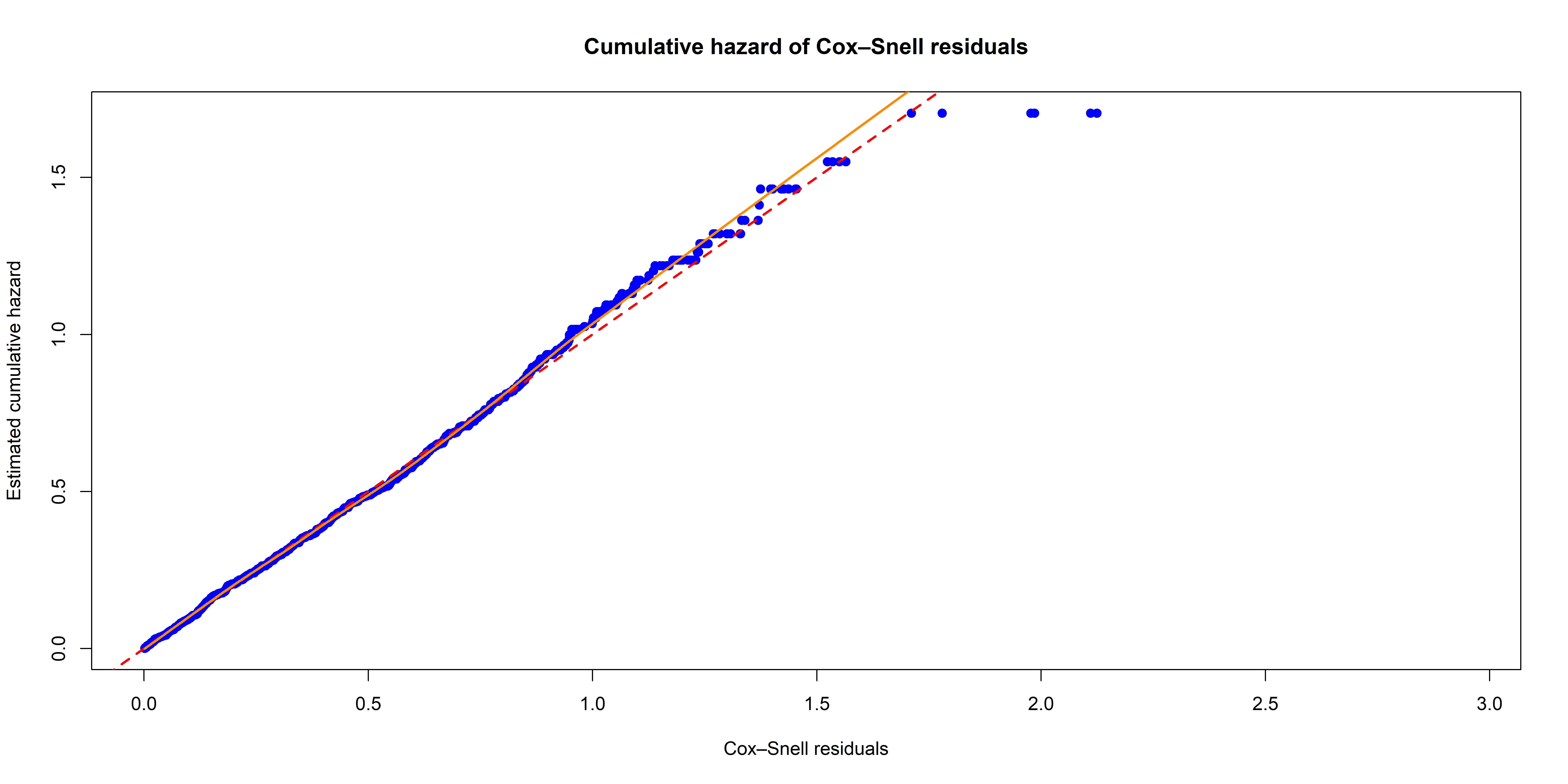}
  \centering
  \renewcommand{\thefigure}{S1}
  \caption{Plot of the estimated cumulative hazard of the Cox–Snell residuals against their values for the adjusted semi-parametric Cox model in the RTOG9202 application.
The red (dashed) line represents the theoretical reference line (y = x) expected under a well-fitting model, while the orange (solid) line shows the smoothed estimate of the data scatter.}
\label{fig1}
\end{figure*}

\clearpage


\begin{figure*}
 \centering
  \includegraphics[width=16.8cm]{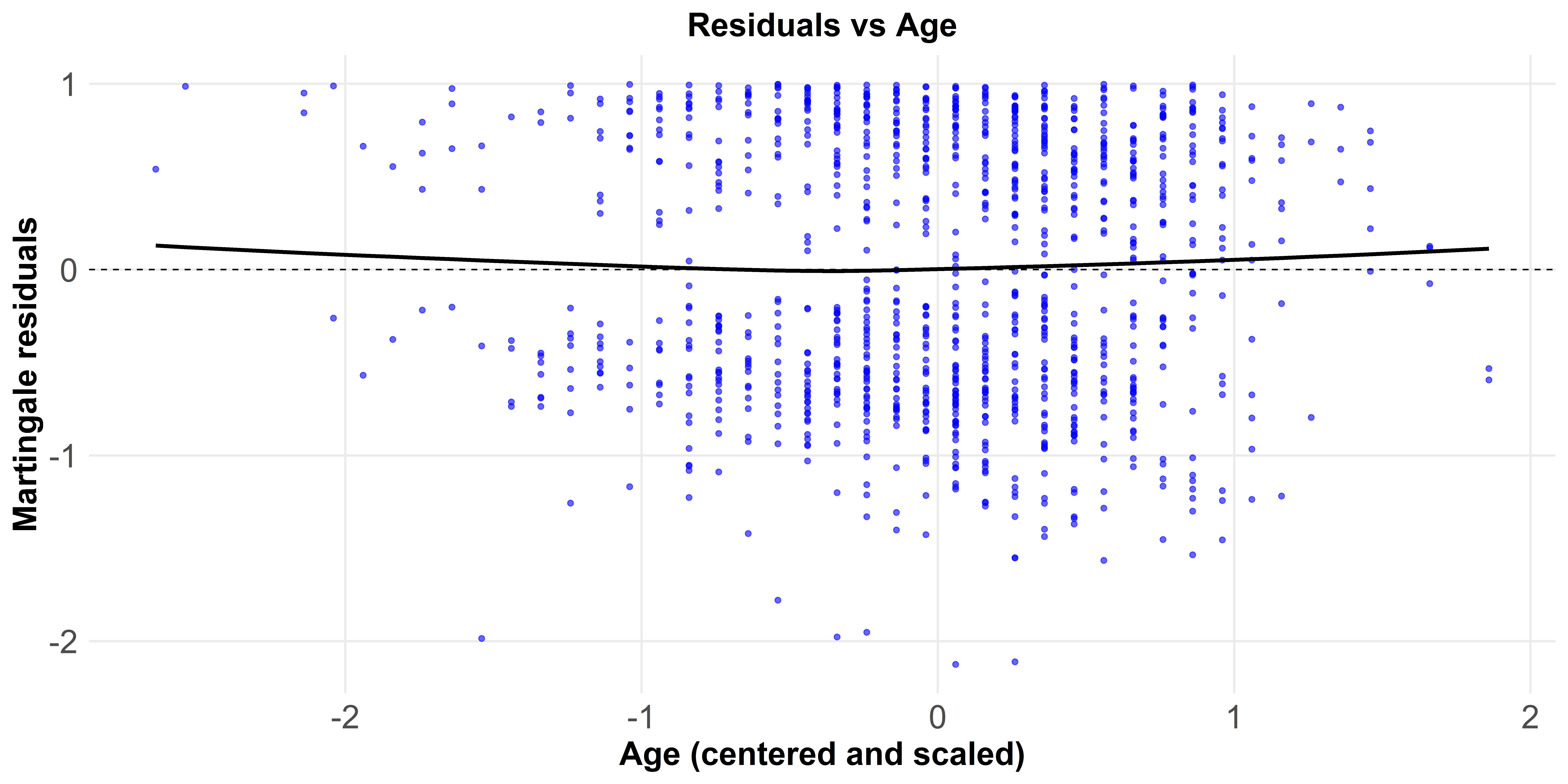}
  \centering
  \renewcommand{\thefigure}{S2}
  \caption{Plot of the martingale residuals against the age covariate (centered and scaled) for the adjusted semi-parametric Cox model in the RTOG9202 application.
The black curve represents the smoothed estimate of the data scatter. }
\label{fig1}
\end{figure*}


\begin{figure*}
 \centering
  \includegraphics[width=16.8cm]{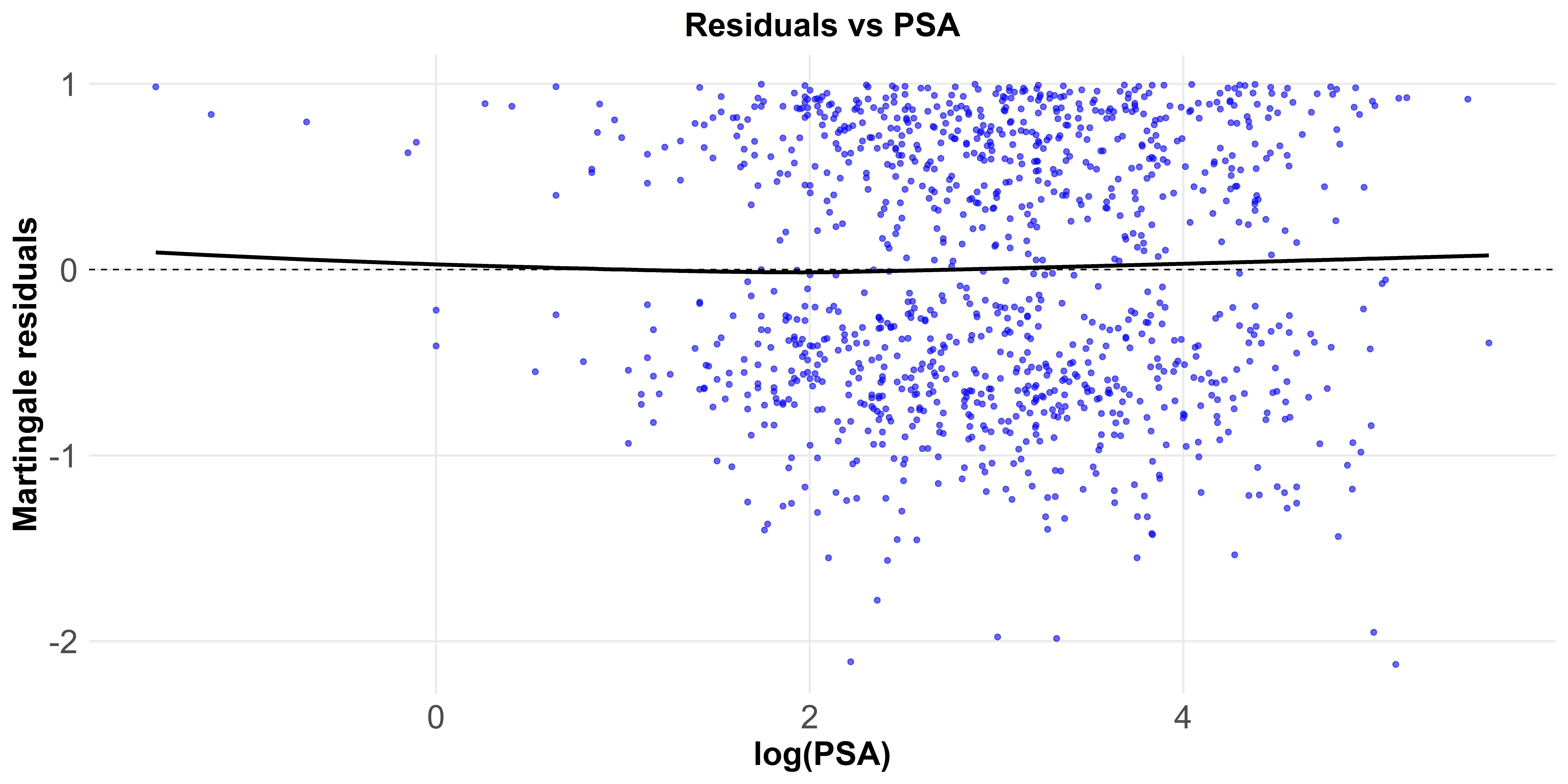}
  \centering
  \renewcommand{\thefigure}{S3}
  \caption{Plot of the martingale residuals against the PSA covariate (log-transformed) for the adjusted semi-parametric Cox model in the RTOG9202 application.
The black curve represents the smoothed estimate of the data scatter.}
\label{fig1}
\end{figure*}

\end{document}